\documentclass[10pt,letterpaper]{article}

\usepackage[top=3cm,left=2.5cm,right=2.5cm,bottom=3cm]{geometry}

\usepackage{parskip}
\usepackage[table]{xcolor}
\usepackage{array}
\usepackage{caption}

\newcolumntype{+}{!{\vrule width 2pt}}

\makeatletter
\renewcommand{\@biblabel}[1]{\quad#1.}
\makeatother

\usepackage{lastpage,fancyhdr,graphicx}
\usepackage{epstopdf}

\fancyheadoffset[L]{2.25in}
\fancyfootoffset[L]{2.25in}
\input commands.cmd

\graphicspath{{./figures/}}

\begin{document} \vspace*{0.2in}


\begin{flushleft}
{\Large \textbf\newline{Information flow in multilayer perceptrons: an in-depth analysis}} \newline 

Giuliano Armano\textsuperscript{*}
\\
\bigskip
Former Associate Professor at the Dept. of Mathematics and Computer Science, University of Cagliari, Cagliari, Italy
\\
\bigskip

* email: armano.giuliano@gmail.com

\end{flushleft}

\section*{Abstract} 
Analysing how information flows along the layers of a multilayer perceptron is a topic of paramount importance in the field of artificial neural networks. After framing the problem from the point of view of information theory, in this position article a specific investigation is conducted on the way information is processed, with particular reference to the requirements imposed by supervised learning. To this end, the concept of information matrix is devised and then used as formal framework for understanding the aetiology of optimisation strategies and for studying the information flow. The underlying research for this article has also produced several key outcomes: i)~the definition of a parametric optimisation strategy, ii)~the finding that the optimisation strategy proposed in the information bottleneck framework shares strong similarities with the one derived from the information matrix, and iii)~the insight that a multilayer perceptron serves as a kind of "adaptor", meant to process the input according to the given objective.


\section{Introduction} \label{sec:introduction} 

There has been a heated debate in the research community about what is meant for ``relevant information'', in particular when the focus is set on machine learning tasks.
%
The definition of relevance (as applied to statistics and information theory) dates back to 1922, with the foundational work of \authors{Fisher}~\cite{bib:fisher:1922}, who studied the concept of sufficient statistic for parametric distributions.
Subsequently, \authors{Lehmann and Scheffé}~\cite{bib:lehmann:1950} introduced the notion of minimal sufficient statistic, showing that there is a minimal partition of the sample space able to capture its relevant components with respect to a given parameter.
%
A bridge between sufficient statistic and information theory was established by \authors{Kullback and Leibler}, with their KL divergence~\cite{bib:kullback:leibler:1951}, which can also measure to what extent a joint distribution of two variables differs from the product of the corresponding marginal distributions (recall that the cited product is a markup for conditional independence).
%
An entropic watershed on this matter was later established by \authors{Tishby et al.}~\cite{bib:tishby:1999}, who claim that information theory provides strong support for identifying the relevant information that algorithms are expected to identify and extract from input sources.
Following this insight, the authors use an information theoretic approach based on rate distortion theory~\cite{bib:davisson:1972} and rooted in the work of Shannon communication theory~\cite{bib:shannon:1948}. Self consistent equations and an iterative procedure for finding useful representations of the input source are also devised therein. A central role in that proposal is played by the concept of information bottleneck (\IB), which is aimed at finding a trade-off between accuracy and complexity (namely, input source compression).
The underlying scenario is described by a Markov chain $\Y \rightarrow \X \rightarrow \T$, which involves two random variables $\X$ and $\Y$, together with an intermediate variable $\T$.
The corresponding interpretation is that \X is generated from \Y (e.g., \X is a noisy version of \Y), whereas \T is a stochastic transformation of \X independent of \Y given \X.

The \IB framework has been framed in a variety of scenarios, including Gaussian variables~\cite{bib:checkik:2005}, continuous variables via variational methods~\cite{bib:alemi:2016}, and geometric clustering~\cite{bib:strouse:2019}.
A specific mention should be made on the application of the \IB in understanding the mechanisms that make deep neural networks so effective.
%
In particular \cite{bib:tishby:2015} and \cite{bib:shwartz-ziv:2017} report an in depth analysis about the learning strategy put into practice by the deep learning technology, in which multiple layers are typically used to progressively extract higher-level features from the input source. 
%
As an alternative to the main stream defined by the \IB principle, in 2016 \authors{Strouse and Schwab} introduce the deterministic \IB (\DIB for short)~\cite{bib:strouse:2016}, claiming that it better captures the notion of compression. The solution to the \DIB problem is a deterministic encoder, as opposed to the stochastic encoder proposed by \authors{Tishby and Zaslavsky}. A comparative benchmarking performed on \IB vs. \DIB shows that the latter outperforms the former in terms of cost function, together with a significant speed-up.

Following the original proposal by \authors{Tishby et al.}, there has been considerable debate regarding the \IB objective function, which characterises the trade-off between compressing the input representation and preserving the information relevant for predicting the output.
With \T intermediate bottleneck variable situated between two random variables \X and \Y, the authors define the following Lagrangian:
\begin{equation*}
  \lagrangian{\cprobability{t}{x}} = \mutual{\X ; \hlayer } - \beta \cdot \mutual{\hlayer ; \Y}
\end{equation*}
According to their analysis, varying the Lagrange multiplier $\beta$ allows to control the trade-off between prediction and compression. 
%
The role of $\beta$ has been criticised by \authors{Wu et al.}~\cite{bib:wu:2019}, who point out that it is typically chosen empirically and that there is also a lack of theoretical understanding about its connection with the concept of learnability, with the intrinsic nature of the dataset and with the capacity of the model. After devising the concept of \IB-Learnability, the authors use it to show that a phase transition occurs (from the inability to the ability to learn), with varying the \IB objective. They also point out that when  $\beta$ is improperly chosen, learning cannot occur. 

\down[0.2] This position article outlines a formal framework for studying multilayer perceptrons (\MLPs) from an information-theoretic perspective. 
To this end, the concepts of mutual information and entropy are used as leading reference tools for investigating to what extent an \MLP is able to preserve relevant information while filtering out the irrelevant one.~The interaction between \emph{relevance} and \emph{filtering ability} coalesces in a $2 \times 2$ \emph{information matrix}, called \IM hereinafter.
Further outcomes of the research from which this article originates are: i)~the definition of a parametric optimisation strategy, ii)~the finding that the optimisation strategy proposed in the \IB framework shares strong similarities with the one derived from the information matrix, and iii)~the insight that an \MLP serves as a kind of "adaptor", meant to process the input according to the given objective.
The remainder of this article is structured as follows:
Section~\ref{sec:materials-and-methods} gives some details about materials and methods, including short summaries on the information processing performed from an unsupervised and a supervised perspective.
Section~\ref{sec:information-matrix} illustrates the conceptual view of \IM, starting from the notions of information relevance and information filtering.
Section~\ref{sec:information-matrix:deterministic} illustrates the deployment of \IM in deterministic scenarios. 
Section~\ref{sec:information-matrix:stochastic} extends the analysis to non-deterministic scenarios. 
 Section~\ref{sec:optimisation-strategies} gives a close look to the aetiology of optimisation strategies, taking \IM as conceptual and formal reference.
Section~\ref{sec:information-flow-in-mlps} deepens the study of information flow in standard \MLPs.
Section~\ref{sec:conclusions} ends the article and gives a sketch about future work.


\section{Materials and Methods} \label{sec:materials-and-methods}

After briefly recalling the definition of mutual information, in this section a qualifying analysis on the kinds of information dealt with in unsupervised and supervised machine learning settings is made. The latter perspective is then selected for further investigation, focusing in particular on the concepts of information relevance and filtering ability.
This section also provides a set of notational conventions for entropy and mutual information, specifically devised to improve the clarity of the presentation.

\subsection{A brief recall about mutual information}

From an information theoretic perspective, the mutual information between two random variables \X and \Y, say \mutual{\X; \Y}, quantifies the amount of information that knowing one variable provides about the other. Notably, mutual information is a sound and unbiased tool for estimating the statistical dependency between two variables, the way used to capture this dependency being agnostic with respect to the underlying probability distributions.
In the cited seminal work dated 1948 (where mutual information was called ``rate of transmission''), Shannon gave its definition first in terms of probability distributions and then as combination of various kinds of entropy.
Few years later, Kullback and Leibler~\cite{bib:kullback:leibler:1951} have shown that, given a joint distribution \joint{\X,\Y}, mutual Information can be expressed as the KL divergence between \joint{\X,\Y} and the product of the corresponding marginals \marginal{\X} and \marginal{\Y}.

For reasons that will be clear afterwards, in this article mutual information is always decomposed in terms of entropies and conditional entropies, according to the following equivalences:
\begin{equation} \label{eq:mutual-information:generic}
\mutual{\X ; \Y} = \HH[\X] - \cHH[\X|\Y] = \HH[\Y] - \cHH[\Y|\X] = \mutual{\Y; \X}
\end{equation}

The equation above makes clear that mutual information is invariant with respect to the ordering between the random variables at hand. In fact, one can think about \mutual{\X; \Y} as the remaining uncertainty about \X once that the conditional dependency between \X and \Y is known, or vice versa.

As preliminary exemption, it is worth pointing out that all random variables mentioned in this article are assumed to be discrete, meaning that if originally continuous they underwent a discretisation process that makes use of a binning and thresholding. This need arises from concurrent causes; in particular i)~under specific conditions mutual information between continuous variables can also be infinite and ii)~dealing with continuous variables makes difficult to estimate mutual information from the available data.
For more information on this matter, see in particular \cite{bib:tishby:1999}, \cite{bib:alomrani:2021} and~\cite{bib:adilova:2023}.

\subsection{Information handling in unsupervised settings}
    
Unsupervised techniques are pervasive with respect to a broad range of research and application fields. Without claiming to be exhaustive, let us recall  denoising in image processing~\cite{bib:ilesamni:2021:review}, autoencoders~\cite{bib:kramer:1991}, principal component analysis~(\PCA)~\cite{bib:jolliffe:2016,bib:sholz:2008:review}, and spectral clustering~\cite{bib:shaham:2018}. 
These application/research fields do not necessarily follow the same rules, as a processing aimed at guaranteeing zero or controlled loss is important in some cases, whereas lossy compression is even favoured in others.

With \X input source and \f transformation applied to it, mutual information allows to establish a bridge between \X and \FX, as follows:
\begin{equation} \label{eq:mutual-info:backward} 
  \mutual{\X ; \FX} = \HH[\FX] - \cHH[\FX|\X] = \HH[\X] - \cHH[\X|\FX]
\end{equation}
Eq.~\ref{eq:mutual-info:backward} can be used to investigate two important aspects of the transformation enacted by \f on the input source \X. Let us examine them separately:

\emph{Stochasticity of \f.} In presence of non-determinism, \f injects noise while processing the input source \X. The conditional entropy \cHH[\FX|X] quantifies the degree to which \f corrupts \X through this noise injection (the higher the value, the noisier the output). A null value of this quantity guarantees determinism.

\emph{Lossless encoding/compression}. Only deterministic transformations that enact lossless encoding/compression allow to preserve the information embedded in the input source \X. The conditional entropy \cHH[\X|\FX] quantifies the degree of information loss, where higher values correspond to greater loss. A null value of this quantity guarantees absence of loss.

\down[0.2] To better highlight the relation that holds between the unsupervised perspective and data source encoding/compression, let us examine two characterising examples, i.e. principal component analysis (controlled loss) and autoencoders (lossless).

\down[0.1] \emph{Principal Component Analysis}. \PCA is a canonical example of transformations that apply controlled loss to the source signal. This dimensionality reduction strategy is primarily motivated by the need to mitigate the curse of dimensionality, a scenario where the number of features rivals or exceeds the number of observations. By projecting the data onto a lower-dimensional space, \PCA is designed to retain most of the input information while drastically reducing the number of features. The key mechanism involves ordering the new features (principal components) according to their ability to preserve the input variance, effectively prioritizing the most informative dimensions.

\down[0.1] \emph{Autoencoders}. Autoencoders can be thought of as engines fully compliant with the encoding-decoding vision borrowed from signal transmission theory. The target for autoencoders is expected to be coincident with (or very similar to) the source data. As a consequence, with \hlayer compressed representation of the input signal \X and \Xstep[out] output of a decoder aimed at reconstructing \X from \hlayer, \mutual{\X ; \hlayer} and \mutual{\hlayer ; \Xstep[out]} show themselves as two sides of the same coin. In fact, the encoder that generates \hlayer is entrusted with enforcing a lossless transformation of \X (or at least a lossy transformation with a controlled loss) and the decoder is designed for reconstructing \X from \hlayer to its maximum extent. Beyond the common inspiring principle, there are several kinds of autoencoders, including Vanilla, denoising, convolutional and variational (the interested reader may consult, for instance, \authors{Li et al.}~\cite{bib:li:2023} for a comprehensive survey on this matter).

\subsection{Information handling in supervised settings}

Supervised learning is characterized not only by an input source \X and a transformation \f, but also by the presence of a target variable \Y.
In this setting \f is a black-box whose aim is to filter in only the information of \X required to predict \Y.
The introduction of a target promotes two distinct perspectives, i.e. backward and forward. In fact, from \F one can look backward to the source or forward to the target.
The dominant measure for looking backward is \mutual{\X;\FX}, which accounts for the amount of information filtered in by \f and made available for prediction (see Eq.~\ref{eq:mutual-info:backward}).
Note that there is no difference between unsupervised and supervised settings when looking backward, as the focus in both cases is to ascertain how much information about \X has been preserved.
Conversely, the dominant measure for looking forward is \mutual{\FX ; \Y}, which accounts for the amount of information that is relevant for predicting \Y.
In particular, the following definition holds:
\down[0.1] \begin{equation} \label{eq:mutual-info:forward}
   \mutual{\FX ; \Y} = \HH[\FX] - \cHH[\FX|\Y] = \HH[\Y] - \cHH[\Y|\FX]
\end{equation}

where \cHH[\FX|\Y] represents the amount of irrelevant information left unremoved by \f and \cHH[\Y|\FX] the amount of relevant information that is missing in \FX.
Naturally, any machine learning algorithm aimed at building a predictive model is expected to minimise the impact of these unwanted aspects, which represent competing objectives that require careful trade-off analysis. In particular, the trivial solution of preserving all information embedded in \X (e.g. by compressing the input source) may not be appropriate for generating effective models. It is true that in so doing no relevant information is lost; however, the generalisation ability of machine learning algorithms is not favoured here, as also the irrelevant information is left unremoved.
A way to escape from this dilemma has been depicted by \authors{Shwartz-Ziv and Tishby}~\cite{bib:shwartz-ziv:2017}, who set the focus on the trade-off between fitting and compression.
In particular, the authors claim that during the training process of an \MLP fitting comes first, followed by compression.
In fact, several studies have criticised this claim.
In particular, \authors{Saxe et al.} observe that compression is not universal, as it depends on which activation function is adopted (e.g., ReLU-based networks may not exhibit compression).
\authors{Golfeld et al.} showed that the compression phase is not consistently observed in practice and may be a consequence of the stochastic search.
\authors{Amjad and Geiger} go further by observing that compression might be an artifact that depends on how mutual information is estimated, rather than being a fundamental property of learning.
Besides the controversial arguments, in any practical setting one should expect  that, due to \f, some irrelevant information is left unremoved, whereas some relevant information is lost (beyond the a-priori lack of information, if any).
A further aspect that needs attention is that in a non-deterministic scenario not all the information filtered in by \f is available for prediction. This can be highlighted by looking from \FX backward to \X, through Eq.~\ref{eq:mutual-info:backward}, which makes clear that \mutual{\X ; \FX} may not be coincident with the information filtered in by \f. The watershed between the two scanarios is measured by \cHH[\FX|\X], which quantifies the stochasticity of \f. 

\subsection{Foremost definitions that apply to supervised learning}

In a supervised setting, the notions of relevant information and filtering ability take on a specific significance: all information useful to predict the target is relevant, the ability to retain it while removing the irrelevant one being a critical issue to be dealt with. 
Due to the specificity of the task, in which one should assume that the information flows from the input source \X throughout a transformation \f (provisionally considered a black-box), the notation for expressing ordinary entropies and mutual information will be customised for the ease of reading. 

\subsubsection{Notational conventions for the relation \handlemath{$\X \rightarrow \Y$}}

\begin{figure}[ht] 
     \center \includegraphics[scale=0.45]{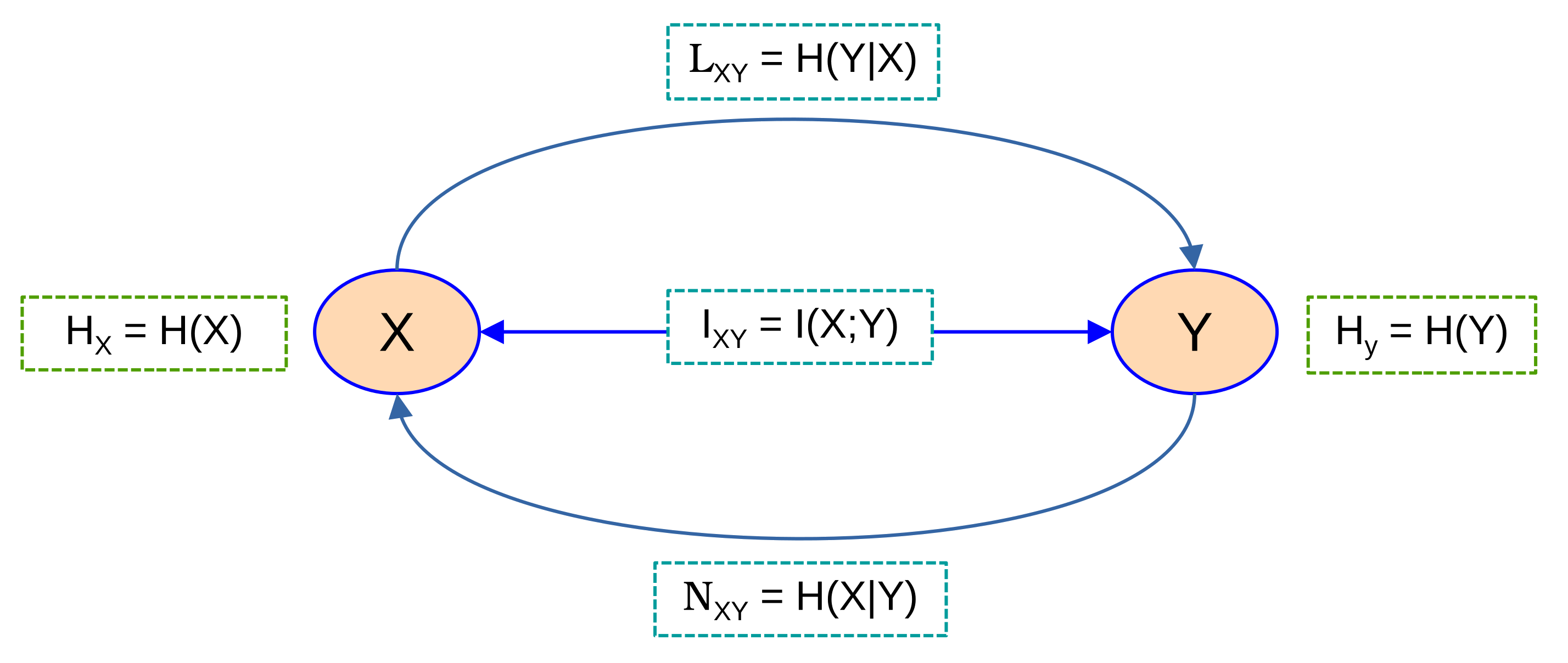}
     \caption{Notational conventions for denoting entropies and mutual information useful for analysing the relation between an input source \X and a target \Y. \noisexy (standing for noise) and \lossxy (standing for lack/loss of information) are both conditional entropies.} \label{fig:mutual-info:xy}
\end{figure}

Figure~\ref{fig:mutual-info:xy} highlights that \mutual{\X ; \Y} leans on entropies and conditional entropies, which in this article will be denoted as follows:
\begin{align} \label{eq:mutual-info:xy}
  \begin{split}
  \Hx \eqdef \HH[\X] \quad \, & \quad \text{Information embedded in \X} \\ 
  \Hy \eqdef \HH[\Y] \quad \, & \quad \text{Information embedded in \Y} \\
  \noisexy \eqdef \cHH[\X|\Y] & \quad \text{Information of \X not relevant for predicting \Y} \\
  \lossxy \eqdef \cHH[\Y|\X] & \quad \text{Lack of information about \Y}
  \end{split}
\end{align}
As for the above redefinitions of conditional entropies, they follow a specific conceptualisation. In particular, \noisexy measures the amount of irrelevant information embedded in \X, to be considered \emph{noise} by any algorithm aimed at finding a predictive model. Conversely, \lossxy measures the lack of relevant information (in a way, an a-priori \emph{loss}), which cannot be recovered by any further processing activity. This impossibility to recover is in accordance with the principle called ``data processing inequality'' (\DPI for short), which asserts that no deterministic or stochastic manipulation of a random variable can provide more information about an original signal than the variable itself contained (see for example \cite{bib:beaudry:2012} on this matter).

Note that there is no ambiguity in associating noise and lack/loss to the corresponding conditional entropies, as the relation $\X \rightarrow \Y$ has a direction.
Based on the above definitions, Eq.~\ref{eq:mutual-information:generic} can be rewritten as follows:
\begin{equation}
     \mutual{\X; \Y} \eqdef \mutualxy  =  \Hx - \noisexy = \Hy - \lossxy 
\end{equation}

\subsubsection{Notational conventions for the relations \handlemath{$\X \rightarrow \FX$} and \handlemath{$\FX \rightarrow \Y$}}

\begin{figure}[ht] 
     \center \includegraphics[scale=0.45]{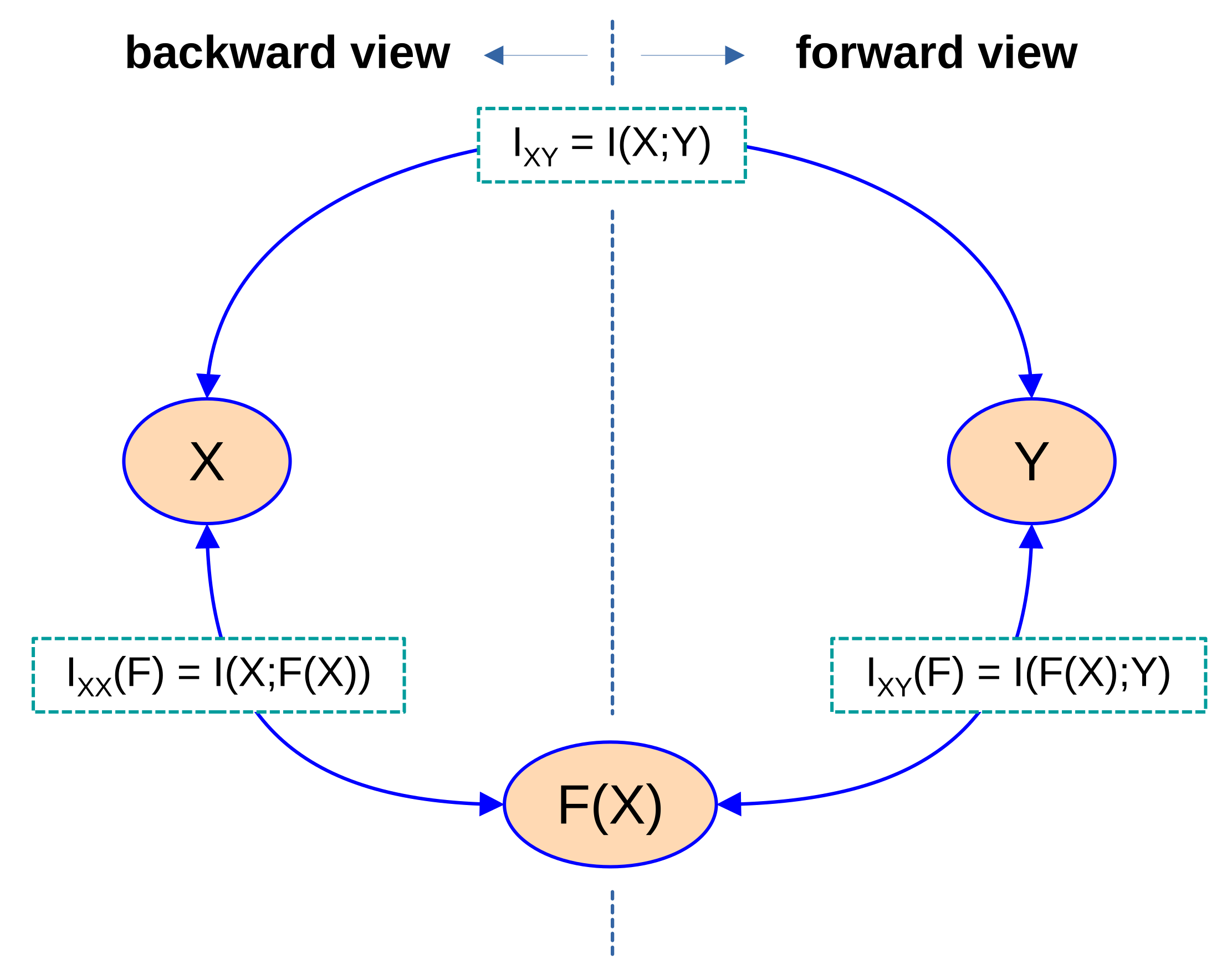}
     \caption{Notational conventions for denoting significant entropies and mutual information useful for analysing the relation between $\X \rightarrow \Y$ throughout a transformation \f that stands in between.} \label{fig:mutual-info:xfy}
\end{figure}

As illustrated in Figure~\ref{fig:mutual-info:xfy}, the transformation \f stands in between \X and \Y. Hence, it represents a sort of ``pivot'' that allows to look backward to the source and forward to the target throughout \mutual{\X ; \FX} and  \mutual{\FX ; \Y}, respectively. Before giving specific notational conventions for these quantities, we need to lean on again on entropies and conditional entropies. In particular, the following definitions hold:
\begin{align}
   \begin{split}
     \Hx[\f] \eqdef \HH[\FX] \quad \, & \quad \text{Information filtered in by} \; \f \\ 
       \noisexx[\f] \eqdef \cHH[\FX|\X] & \quad \text{Noise due to the stochasticity of \f} \\ 
      \lossxx[\f] \eqdef \cHH[\X|\FX] & \quad \text{Loss of information about \X due to \f} \\
      \noisexy[\f] \eqdef \cHH[\FX|\Y] & \quad \text{Irrelevant information of \X left unremoved by \f} \\ 
      \lossxy[\f] \eqdef \cHH[\Y|\FX] & \quad \text{Lack of relevant information + loss due to \f}
   \end{split}
\end{align}

Also in this case the redefinitions of conditional entropies follow a specific conceptualisation. Let us consider backward and forward perspectives separately. 

\down[0.2] \emph{Backward perspective} -- \noisexx[\f] denotes the uncertainty about \X due to the stochastic behaviour of \f; hence, this term is null only when \f is deterministic. Conversely, \lossxx[\f] represents the loss of information about \X due to \f; hence this term is null only when \f enacts a lossless encoding or compression (however, in a non-deterministic scenario only \emph{near-lossless} encoding/compression can occur).

\down[0.2] \emph{Forward perspective} -- \noisexy[\f] denotes the information of \X not relevant for predicting \Y and left unremoved by \f. Conversely, \lossxy[\f] represents the relevant information that is missing in \X, plus the one that has been lost by \f. 

\down[0.3] Summarising, the following definitions hold for the ``backward'' and ``forward'' kinds of mutual information:
\begin{align} \label{eq:rewriting:Ixx(f)-vs-Ixy(f):with-noise-and-loss}
  \begin{split}
    \mutual{\X ; \FX} \eqdef \mutualxx[\f] &= \Hx[\f] - \noisexx[\f] = \Hx - \lossxx[\f] \\
    \mutual{\FX ; \Y} \eqdef \mutualxy[\f] &= \Hx[\f] - \noisexy[\f] = \Hy - \lossxy[\f]
  \end{split}
\end{align}

A further definition follows, deemed useful to clarify a critical point related to the distinction between lack of information and information loss. As pointed out, \lossxy[\f] includes both the a-priori lack of relevant information and the amount of relevant information lost by \f. As the latter will be useful afterwards, let us give a specific definition for it:
\begin{equation*}
     \dlossxy[\f] \eqdef \lossxy[\f] - \lossxy \ge 0
\end{equation*}
Note that the inequality $\dlossxy[\f] \ge 0$ points out once again that the a-priori lack of information cannot be recovered by any further processing.


\section{Devising the information matrix} \label{sec:information-matrix}

In this section, the information matrix will be first illustrated from a conceptual perspective and then represented in terms of mutual information and entropies.
Afterwards, \IM will be taken as sound starting point for studying the aetiology of optimisation strategies and for analysing the information flow across the layers of an \MLP.

Note that the content of this section is not specific to \MLPs, being applicable to any supervised learning scenario where a function \f must be optimized to predict a target \Y from an input source \X.
To emphasize the role of \f as a pivot that stands in between \X and \Y, the notation \IM[\f] will be used hereinafter.

\subsection{\handlemath{\IM[\f]} from a conceptual perspective}

In a supervised setting, the ``information game'' is played by three main actors: the input source \X, the transformation \f, and the target \Y.
In this analysis the most general assumptions are made on the underlying process, namely: i)~\f may be stochastic, ii)~\X may not contain all the information needed to predict \Y, iii)~part of the irrelevant information may be left unremoved by \f, and iv)~part of the relevant information may be lost due to \f.

Characterising the interlacing between information relevance and filtering ability is a step that leads to the definition of \IM[\f].
Relevance is focused on the fact that typically only part of the information embedded in \X is actually required to predict \Y. Hence, given \X and \Y, a clear separation holds between relevant and irrelevant information. Conversely, the filtering ability regards to what extent \f is able to retain relevant information while removing the irrelevant one. Note that the former is a property of $\X \rightarrow \Y$, whereas the latter is a property of $\FX \rightarrow \Y$.
Table~\ref{tab:im:conceptual-view} highlights the semantics of the crossings between relevance and filtering ability. The components of \IM[\f] are marked using provisional symbols, i.e. \ca, \cb, \cc and \cd, together with a label that summarises the relating kind of information.

Notably, the terms reported in Table~\ref{tab:im:conceptual-view} will be employed throughout this article with consistently-defined semantics. In particular, i)~filtered in/filtered out are only concerned with the filtering process enacted by \f, regardless of the kind of information being dealt with, ii)~removed/left unremoved are used in association with irrelevant information, whereas iii)~retained/lost are used in association with relevant information.
\begin{table}[!ht]
\centering \small \renewcommand{\arraystretch}{2.0}
\begin{tabular}{ l l l }
           & Filtered out by {\f} & Filtered in by {\f} \\
 \cline{2-3}
  Not relevant for \Y & \ca = Removed & \cb = Left unremoved \\ 
  Relevant for \Y & \cc = Lost  &  \cd = Retained \\
  \cline{2-3} \\
\end{tabular}
\caption{How the information embedded in an input source \X is fragmented according to relevance and filtering ability. The semantics of each item in the table is given by the crossing between the corresponding row and column. As arbitrary choice, relevance pertains to rows and filtering ability to columns.} \label{tab:im:conceptual-view}
\normalsize
\end{table}

With \mutualxy representing the relevant information embedded in \X and \Hx[\f] the information filtered in by \f, the following notational conventions are obtained by taking the complement of the given quantities. In symbols:
\begin{align*}
   \nmutualxy &\eqdef \Hx - \mutualxy \equiv \noisexy \quad \quad \text{Irrelevant information}\\
   \Hx[\nf] &\eqdef \Hx - \Hx[\f] \quad \quad \quad  \phantom{123} \text{Information filtered out by \f}
\end{align*}

The intended semantics of the above definitions is clear: an overline above \mutualxy denotes a complement on the irrelevant side, whereas an overline above \f denotes a complement on the filtered-out side.
The same semantics can be adopted to make explicit the content of \IM[\f], as follows: 
\begin{align*} 
    \text{\ca} \; &:: \; \nmutualxy[\nf] \quad \quad \text{Irrelevant information that has been \emph{removed}} \\
    \text{\cb} \; &:: \; \nmutualxy[\f] \quad \quad \text{Irrelevant information that has been \emph{left unremoved}} \\
    \text{\cc} \; &:: \; \mutualxy[\nf] \quad \quad \text{Relevant information that has been \emph{lost}} \\
   \text{\cd} \; &:: \; \mutualxy[\f] \quad \quad \text{Relevant information that has been \emph{retained}}
\end{align*}

Note that, as the overline on mutual information pertains to rows whereas the one on \f pertains to columns, there is no ambiguity at jointly using them as done with \nmutualxy[\nf].

\subsection{Constraints that apply to the information matrix}

Notwithstanding the fact that the it will be defined in terms of entropies and mutual information, \IM[\f] belongs to the broad category of binary confusion matrices. Hence, it should not be surprising that its degrees of freedom are limited by some constraints that depend on the way it has been conceived. In fact, irrelevant and relevant information amount to \noisexy and \mutualxy, regardless of the filtering activity performed by \f.
Table~\ref{tab:im:constraints} displays the above fundamental constraints and other relevant constraints that complete the analytical framework. For the sake of readability they have been grouped according to their pertinence to rows or columns.

\begin{table}[!ht]
\centering \small \renewcommand{\arraystretch}{1.9}
\begin{tabular}{ l l l }
 \cline{2-3}
     By row & $\nmutualxy + \mutualxy = \Hx$ & Overall information \\ 
                   & $\nmutualxy[\nf] + \nmutualxy[\f] = \nmutualxy = \noisexy$ & Irrelevant information \\ 
                  & $\mutualxy[\nf] + \mutualxy[\f] = \mutualxy$ & Relevant information  \\ 
  \cline{2-3}
     By column &  $\Hx[\nf] + \Hx[\f] = \Hx$ & Overall information \\ 
                 & $\nmutualxy[\nf] + \mutualxy[\nf] = \Hx[\nf]$ & Information  filtered out \\ 
                 & $\nmutualxy[\f] + \mutualxy[\f] = \Hx[\f]$ & Information  filtered in \\ \\
\end{tabular}
\caption{Summary on the constraints that hold for items relating to \IM[\f], together with their semantics. The constraints are grouped according to their pertinence to rows or columns.} \label{tab:im:constraints}
\normalsize
\end{table}

\subsection{\handlemath{\IM[\f]} given in terms of entropy and mutual information}

According to Table~\ref{tab:im:constraints}, representing \nmutualxy[\nf], \nmutualxy[\f], \mutualxy[\nf] and \mutualxy[\f] is now straightforward. \down[0.15]
\begin{align*}
    \text{\cd} &:: \mutualxy[\f] = \Hy - \lossxy[\f] = \Hy - \lossxy - \dlossxy[\f] = \mutualxy - \dlossxy[\f] \\
    \text{\cc} &:: \mutualxy[\nf] = \mutualxy - \text{\cd} = \mutualxy - \mutualxy[\f] = \lossxy[\f] - \lossxy = \dlossxy[\f] \\
    \text{\cb} &:: \mutualxy[\nf] = \Hx[\f] - \text{\cd} = \Hx[\f] - \mutualxy[\f] = \noisexy[\f] \\
    \text{\ca} &:: \nmutualxy - \text{\cb} = \Hx - \mutualxy - \noisexy[\f] = \noisexy - \noisexy[\f]
\end{align*} 

\begin{table}[!ht]
\centering \small \renewcommand{\arraystretch}{2.0}
\begin{tabular}{ l l l }
           & Filtered out by {\f} $\equiv \Hx[\nf]$ & Filtered in by {\f}  $\equiv \Hx[\f]$ \\
 \cline{2-3}
  Not relevant for \Y  $\equiv \noisexy$ & $\nmutualxy[\nf] = \noisexy - \noisexy[\f]$ & $\nmutualxy[\f] \equiv \noisexy[\f]$ \phantom{-----}\\ 
  Relevant for \Y  $\equiv \mutualxy$ & $\mutualxy[\nf] = \dlossxy[\f]$ &  $\mutualxy[\f] = \mutualxy - \dlossxy[\f]$  \\
  \cline{2-3} \\
\end{tabular}
\caption{How the information embedded in an input source \X is fragmented in the general case, according to its relevance with the target \Y and to the transformation \f. All terms are made explicit using the notational conventions that hold for entropies and conditional entropies.} \label{tab:im:summary}
\normalsize
\end{table}

\begin{figure}[!ht]
  \centering
  \includegraphics[width=0.60\textwidth]{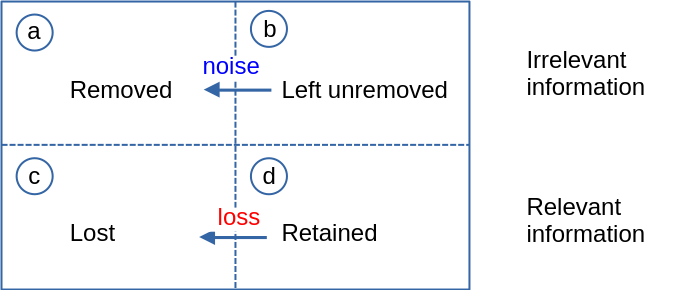}
  \vspace{5mm}
  \caption{Underlying dynamics of \IM[\f], in which both noise and loss flow from right to left, the former at the irrelevant side and the latter at the relevant side. Note that, in principle, the amount of noise and loss may also depend on a non-deterministic behaviour of \f.} \label{fig:im:noise-loss-dynamics}
\end{figure}

\down[0.2] Table~\ref{tab:im:summary} provides a summary of \IM[\f], given in terms of entropies and mutual information. The table also reports the sums over rows and columns, namely \noisexy for the irrelevant side, \mutualxy for the relevant side, \Hx[\nf] for the information filtered out and \Hx[\f] for the information filtered in. Note that a fundamental equation is also embedded in column 2, which allows to move from the forward to the backward view and vice versa. In symbols:
\begin{align} \label{eq:im:stochastic:fundamental-equation}
   \mutualxy[\f] + \noisexy[\f] = \Hx[\f] = \mutualxx[\f] + \noisexx[\f]
\end{align}
Note also that \IM[\f] allows to highlight the pathways followed by the desirable removal of irrelevant information and by the detrimental loss of relevant information. In particular, as pointed out by Figure~\ref{fig:im:noise-loss-dynamics}, the noise is removed along the path \cb $\rightarrow$ \ca, whereas the additional loss follows the path \cd $\rightarrow$ \cc.
Hence, \IM[\f] is in fact a \emph{noise vs. loss matrix}, its content being deeply affected by the actual values of \noisexy[\f] and \lossxy[\f]. 

\down[0.4] The next section contains the analysis of \IM[\f] under the simplifying hypothesis that \f is deterministic. This analysis plays the role of premise, useful to softly introduce concepts and definitions that will be resumed in a more complex form when the non-deterministic scenario will be analysed.


\section{Information matrix for deterministic scenarios} \label{sec:information-matrix:deterministic}

As pointed out, \f is a deterministic transformation only when \noisexx[\f] is null. This assumption has important consequences on various aspects, including i)~the relation between \mutualxy[\f] and \mutualxx[\f], ii)~the ranges that apply to the components of \IM[\f], iii) the characteristics of well-known behavioural patterns, iv)~the representation of noise and loss in a cartesian space, and v)~the isometrics of \mutualxx[\f].

\subsection{Relation between \handlemath{\mutualxy[\f]} and \handlemath{\mutualxx[\f]}}  \label{sub:im:deterministic:ixyf-vs-ixxf}

Under the hypothesis of determinism $\noisexx[\f] = 0$, hence $\Hx[\f] \equiv \mutualxx[\f]$. This equivalence allows to simplify Eq.~\ref{eq:im:stochastic:fundamental-equation} as follows:
\begin{equation} \label{eq:im:deterministic:fundamental-equation}
    \mutualxy[\f] + \noisexy[\f] = \Hx[\f] = \mutualxx[\f]
\end{equation}
This equation, which lies behind deterministic scenarios, clearly highlights that \emph{all} the information filtered in by \f is available for prediction, although subject to the split into relevant and irrelevant. Note that Eq.~\ref{eq:im:deterministic:fundamental-equation} implies the inequality $ \mutualxy[\f] \le \mutualxx[\f]$.

\subsection{Admissible ranges that apply to the elements of \handlemath{\IM[\f]}}  \label{sub:im:deterministic:ranges}

With the goal of finding admissible ranges that hold for the elements of \IM[\f], it is useful to start by studying \noisexy[\f] and \lossxy[\f]. With a deterministic \f, the following constraints hold:
\begin{align}
    \begin{split}
        0 &\le \, \noisexy[\f] \le \noisexy \\
        \lossxy &\le \, \lossxy[\f] \, \le \Hy
    \end{split}
\end{align}

While giving a motivation for maximum values would be trivial, minimum values deserve a comment. Being \f deterministic, in principle it may be able to remove all unwanted noise; hence $min \; \noisexy[\f]=0$. Conversely, $min \; \lossxy[\f] = \lossxy$ as (due to the \DPI) no further processing can contrast the lack of information that holds for the relation $\X \rightarrow \Y$. Notably, $min \; \lossxy[\f] \equiv 0$ signals that all information required to predict \Y is embedded in \X.

\down[0.2] With the ranges of \noisexy[\f] and \lossxy[\f] determined, the element-wise ranges of \IM[\f] can be readily computed. Table~\ref{tab:im:deterministic:ranges} contains a summary information about them.
\begin{table}[!ht]
\centering \small \renewcommand{\arraystretch}{1.5}
\begin{tabular}{ l c c }
           & Min & Max \\
 \cline{2-3}
  \ca \phantom{:} :: $\; \nmutualxy[\nf] = \noisexy - \noisexy[\f]$ \phantom{abc} & 0 & \noisexy\\ 
  \cb \phantom{:} :: $\; \nmutualxy[\f] = \noisexy[\f]$ & 0 & \noisexy\\ 
  \cc \phantom{:} :: $\; \mutualxy[\nf] = \lossxy[\f] - \lossxy$ & 0 & \mutualxy\\ 
  \cd \phantom{:} :: $\; \mutualxy[\f] = \Hy - \lossxy[\f]$ & 0 & \mutualxy \\ \\
\end{tabular}
\caption{Minimum and maximum values for each component of \IM[\f], under the hypothesis of deterministic \f.}\label{tab:im:deterministic:ranges}
\normalsize
\end{table}

\subsection{Behavioural patterns for deterministic scenarios}  \label{sub:im:deterministic:patterns}

\down[0.2] Having established the semantics of \IM[\f], let us put this knowledge into practice by analysing some behavioural patterns that are well-known in the machine learning community:
\begin{itemize}
  \item[--] Pattern \cone$\;\,$:: Lossless encoding / compression\\
  \f filters in all information from \X, including the irrelevant one;
  \item[--] Pattern \ctwo$\,$ :: Maximum discriminative capability \\
  \f removes all irrelevant information while retaining all the relevant one;
  \item[--] Pattern \cthree$\,$ :: Dummy prediction\\
  \f removes all irrelevant information while losing all the relevant one;
  \item[--] Pattern \cfour$\,$ :: Random prediction\\
  \f retains all irrelevant information while losing all the relevant one.
\end{itemize}

\subsubsection{Pattern \cone (lossless encoding / compression)}

In this case all information is filtered in by \f. In symbols, $\noisexy[\f] \equiv \noisexy$ (maximum noise) and $\lossxy[\f] \equiv \lossxy$ (minimum loss). As a consequence, the elements of the first column are null (see Table~\ref{tab:im:deterministic:lossless}). 
To show that this pattern is compliant with the one regarding the general case, let us substitute the values of noise and loss in \IM[\f]:
\begin{align*}
     \text{\ca \, :: \;} & \nmutualxy[\nf] = \noisexy - \noisexy[\f] = \noisexy - \noisexy = 0\\
     \text{\cb \, :: \;} & \nmutualxy[\f] = \noisexy[\f] = \noisexy \\
     \text{\cc \, :: \;} & \mutualxy[\nf] = \dlossxy[\f] = \lossxy - \lossxy = 0\\
    \text{\cd \, :: \;} & \mutualxy[\f] = \mutualxy - \dlossxy[\f] = \mutualxy
\end{align*}
\begin{table}[!ht]
\centering \small \renewcommand{\arraystretch}{1.5}
\begin{tabular}{ l c c }
           & Filtered out by {\f} & Filtered in by {\f} \\
 \cline{2-3}
  Not relevant for \Y & 0 & \noisexy\\ 
  Relevant for \Y & 0 &  \mutualxy  \\
  \cline{2-3} \\
\end{tabular}
\caption{How the information embedded in an input source \X is fragmented under the assumption of lossless encoding / compression. In this case the first column of \IM[\f] is null, meaning that no information whatsoever has been removed, whereas the second column contains all information.} \label{tab:im:deterministic:lossless}
\normalsize
\end{table}

In summary, the assumption that \f enacts a lossless encoding / compression preserves all information embedded in \X. This pattern should be taken as reference for all variants in which a small part of irrelevant information is removed and a small part of relevant information is lost.

Note that, from a theoretical perspective, lossless encoding / compression is not expected to help building the predictive model in itself. However, in practice, compressing the input without loss might help supervised learning algorithms when the number of features is on the order of (or larger than) the number of observations.
Notwithstanding this practical aspect, the typical outcome of lossless encoding / compression is ``rote learning'', meaning that no generalisation is expected to occur while building the predictive model. Recall that, in the artificial neural network community, a lack of generalisation is typically signalled by the unwanted training outcome called overfitting. 

\subsubsection{Pattern \ctwo (maximum discriminative capability)}

This pattern in centred on the assumption that \f achieves the highest level of discriminative capability, meaning that all relevant information is retained and that all the irrelevant one is removed.
The corresponding  \IM[\f] is reported in Table~\ref{tab:im:deterministic:max-discriminative-ability} (note that no hypothesis is made on \lossxy, as in principle there might be a lack of information in \X that prevents the target to be perfectly predicted). 
\begin{table}[!ht]
\centering \small \renewcommand{\arraystretch}{1.5}
\begin{tabular}{ l c c }
           & Filtered out by {\f} & Filtered in by {\f} \\
 \cline{2-3}
  Not relevant for \Y & \noisexy & 0 \\ 
  Relevant for \Y & 0 &  \mutualxy  \\
  \cline{2-3} \\
\end{tabular}
\caption{How the information embedded in a source \X is fragmented in the ideal case in which \f allows to achieve the highest level of discriminative capability.} \label{tab:im:deterministic:max-discriminative-ability}
\normalsize
\end{table}
To show that this pattern is compliant with the one of the general case, let us impose the constraints $\noisexy[\f]=0$ (no noise) and $\lossxy[\f] = \lossxy$ (minimum loss):
\begin{align*}
     \text{\ca \, :: \;} & \nmutualxy[\nf] = \noisexy - \noisexy[\f] = \noisexy - 0 = \noisexy\\
     \text{\cb \, :: \;} & \nmutualxy[\f] = \noisexy[\f] = 0 \\
     \text{\cc \, :: \;} & \mutualxy[\nf] = \dlossxy[\f] = \lossxy - \lossxy = 0 \\
    \text{\cd \, :: \;} & \mutualxy[\f] = \mutualxy - \dlossxy[\f] = \mutualxy - 0 = \mutualxy \\
 \end{align*}

An important subcase occurs when $\lossxy=0$, meaning that \X contains all information required for predicting \Y.  In principle, this subcase allows a trained model to perform as an oracle. The corresponding \IM[\f], derived from the one depicted in Table~\ref{tab:im:deterministic:max-discriminative-ability}, is reported in Table~\ref{tab:im:deterministic:oracle}. 
\begin{table}[!ht]
\centering \small \renewcommand{\arraystretch}{1.5}
\begin{tabular}{ l c c }
           & Filtered out by {\f} & Filtered in by {\f} \\
 \cline{2-3}
  Not relevant for \Y & \noisexy & 0 \\ 
  Relevant for \Y & $0$ &  $\Hy$  \\
  \cline{2-3} \\
\end{tabular}
\caption{How the information embedded in an input source \X is fragmented when an oracle-like behaviour holds. The underlying hypotheses are: i)~no lack of information in \X, ii)~all irrelevant information removed and iii)~all relevant information retained. Note that in this case the overall information \Hx amounts to $\Hy + \noisexy$, since $\lossxy=0$.} \label{tab:im:deterministic:oracle}
\normalsize
\end{table}

In summary, the pattern of maximum discriminative capability is characterised by the removal of all irrelevant information, concurrently with the retention of all the relevant one. This pattern should be taken as reference for all variants in which a small part of irrelevant information is left unremoved and a small part of relevant information is lost. 

Note that, in principle, both pattern \ctwo and pattern \cone allow to generate machine learning models that achieve maximum discriminative capability. However, due to the typical caveats of machine learning algorithms, the removal of irrelevant information would always be preferable.

As for the possibility of reaching an oracle-like behaviour, it is characterised by the additional hypothesis that all information required to predict \Y is contained in \X.  

\subsubsection{Pattern \cthree (dummy prediction)}

This pattern holds when \f acts as ``global shadower'', meaning that both kinds of information are filtered out.
This assumption leads to the \IM[\f] reported in Table~\ref{tab:im:deterministic:dummy-predictor}, which highlights that $\noisexy[\f] = 0$ (i.e. all irrelevant information removed) and $\lossxy[\f] \equiv \Hy$ (i.e. all relevant information lost).
\begin{table}[!ht]
\centering \small \renewcommand{\arraystretch}{1.5}
\begin{tabular}{ l c c }
           & Filtered out by {\f} & Filtered in by {\f} \\
 \cline{2-3}
  Not relevant for \Y & \noisexy & 0 \\ 
  Relevant for \Y & \mutualxy &  0  \\
  \cline{2-3} \\
\end{tabular}
\caption{How the information embedded in an input source \X is fragmented in the event that all information has been shadowed.} \label{tab:im:deterministic:dummy-predictor}
\normalsize
\end{table}
To show that this pattern is compliant with the one regarding the general case, let us derive it by imposing $\noisexy[\f] = 0$ and $\lossxy[\f] = \Hy$:
\begin{align*}
     \text{\ca \, :: \;} &  \nmutualxy[\nf] = \noisexy - \noisexy[\f] = \noisexy - 0 = \noisexy\\
      \text{\cb \, :: \;} &  \nmutualxy[\f] = \noisexy[\f] = 0 \\
     \text{\cc \, :: \;} & \mutualxy[\nf] = \dlossxy[\f] = \Hy - \lossxy = \mutualxy\\
      \text{\cd \, :: \;} & \mutualxy[\f] = \mutualxy - \dlossxy[\f] = \mutualxy - \mutualxy = 0 \\
\end{align*}

In summary, when \f follows this pattern, all information is filtered out, leading to a situation in which no information whatsoever is made available for prediction. This pattern should be taken as reference for all variants in which a small part of irrelevant information is left unremoved and a small part of relevant information is retained.

Note that the most likely outcome of model building with all information filtered out by \f yields a predictor that emits a constant, regardless of the input source.

\subsubsection{Pattern \cfour (random prediction)}

This pattern holds when all relevant information is lost, whereas all the irrelevant one is left unremoved (see Table~\ref{tab:im:deterministic:random-predictor}).
\begin{table}[!ht]
\centering \small \renewcommand{\arraystretch}{1.5}
\begin{tabular}{ l c c }
           & Filtered out by {\f} & Filtered in by {\f} \\
 \cline{2-3}
  Not relevant for \Y & 0 & \noisexy \\ 
  Relevant for \Y & \mutualxy &  0  \\
  \cline{2-3} \\
\end{tabular}
\caption{How the information embedded in an input source \X is fragmented in case of complete loss of relevant information, together with the complete retention of the irrelevant one. Note that, due to the loss:  $\mutualxy[\nf] = \Hy - \lossxy \equiv \mutualxy$.} \label{tab:im:deterministic:random-predictor}
\normalsize
\end{table}
Whilst sharing commonalities with the reference pattern~\cthree, here the emphasis should be put on the duality that holds between random prediction and maximum discriminative capability. 

In summary, when \IM[\f] follows this pattern, the information processing is reversed with respect to the aim of predicting \Y. This pattern should be taken as reference for all variants in which a small part of irrelevant information is removed and a small part of relevant information is retained.

Note that the most likely outcome of model building with such a kind of ``information reversal'' yields a predictor that emits random labels, regardless of the input source.

\subsection{Devising noise-loss diagrams}

On one hand, as noise and loss are central aspects for studying the behaviour of \f, a Cartesian representation aimed at highlighting the process of information transfer should be centred on this couple of measures. On the other hand, the most likely candidates to become axes of this Cartesian space are the elements of the primary diagonal of \IM[\f], namely \nmutualxy[\nf] and \mutualxy[\f]. Fortunately, these two aspects are not in contrast, as \nmutualxy[\nf] is strictly linked to the irrelevant information left unremoved, whereas \mutualxy[\f] is strictly linked to the relevant information lost. In symbols:
\begin{align}
  \begin{split}
    \nmutualxy[\nf] &= \noisexy - \noisexy[\f] \quad \quad \quad \quad \quad \quad 0 \, \le \nmutualxy[\nf] \le \noisexy \\
    \mutualxy[\f] &= \mutualxy - \dlossxy[\f] \quad \quad \quad \quad \quad \; \; 0 \, \le \mutualxy[\f] \; \le \mutualxy \\
  \end{split}
\end{align}
One may wonder where the behavioural patterns previously analysed are located in this hypothetical \emph{noise-loss diagram}.

\down[0.2] \begin{itemize}

  \item[\cone] Lossless encoding / compression coordinates: \pair{0, \mutualxy} \\
       With this pattern $\noisexy[\f] = \noisexy$ and $\lossxy[\f] = \lossxy$; hence: \\
       $\nmutualxy[\nf] = \noisexy - \noisexy[\f] = \noisexy - \noisexy = 0$ \\
       $\mutualxy[\f]=\mutualxy - \dlossxy[\f] = \mutualxy - \big( \lossxy - \lossxy \big) = \mutualxy$
  
  \item[\ctwo] Maximum discriminative capability  coordinates: \pair{\noisexy,\mutualxy} \\
      With this pattern: $\noisexy[\f] = 0$ and $\lossxy[\f] = \lossxy$; hence: \\
      $\nmutualxy[\nf] = \noisexy - \noisexy[\f] = \noisexy - 0 = \noisexy$ \\
      $\mutualxy[\f]=\mutualxy - \dlossxy[\f] = \mutualxy - \big( \lossxy - \lossxy \big) = \mutualxy$
  
  \item[\cthree] Dummy prediction coordinates: \pair{\noisexy,0} \\
       With this pattern $\noisexy[\f] = 0$ and $\lossxy[\f]=\Hy$; hence:\\
       $\nmutualxy[\nf] = \noisexy - \noisexy[\f] = \noisexy - 0 = \noisexy$ \\
       $\mutualxy[\f]=\mutualxy - \dlossxy[\f] = \mutualxy - \big( \Hy - \lossxy \big) = 0$
 
  \item[\cfour] Random prediction coordinates: \pair{0,0} \\
       With this pattern $\noisexy[\f] = \noisexy$ and $\lossxy[\f]=\Hy$; hence: \\
       $\nmutualxy[\nf] = \noisexy - \noisexy[\f] = \noisexy - \noisexy = 0$ \\
       $\mutualxy[\f]=\mutualxy - \dlossxy[\f] = \mutualxy - \big( \Hy - \lossxy \big) = 0$
  
\end{itemize}

\begin{figure}[!ht]
  \centering
  \includegraphics[width=0.90\textwidth]{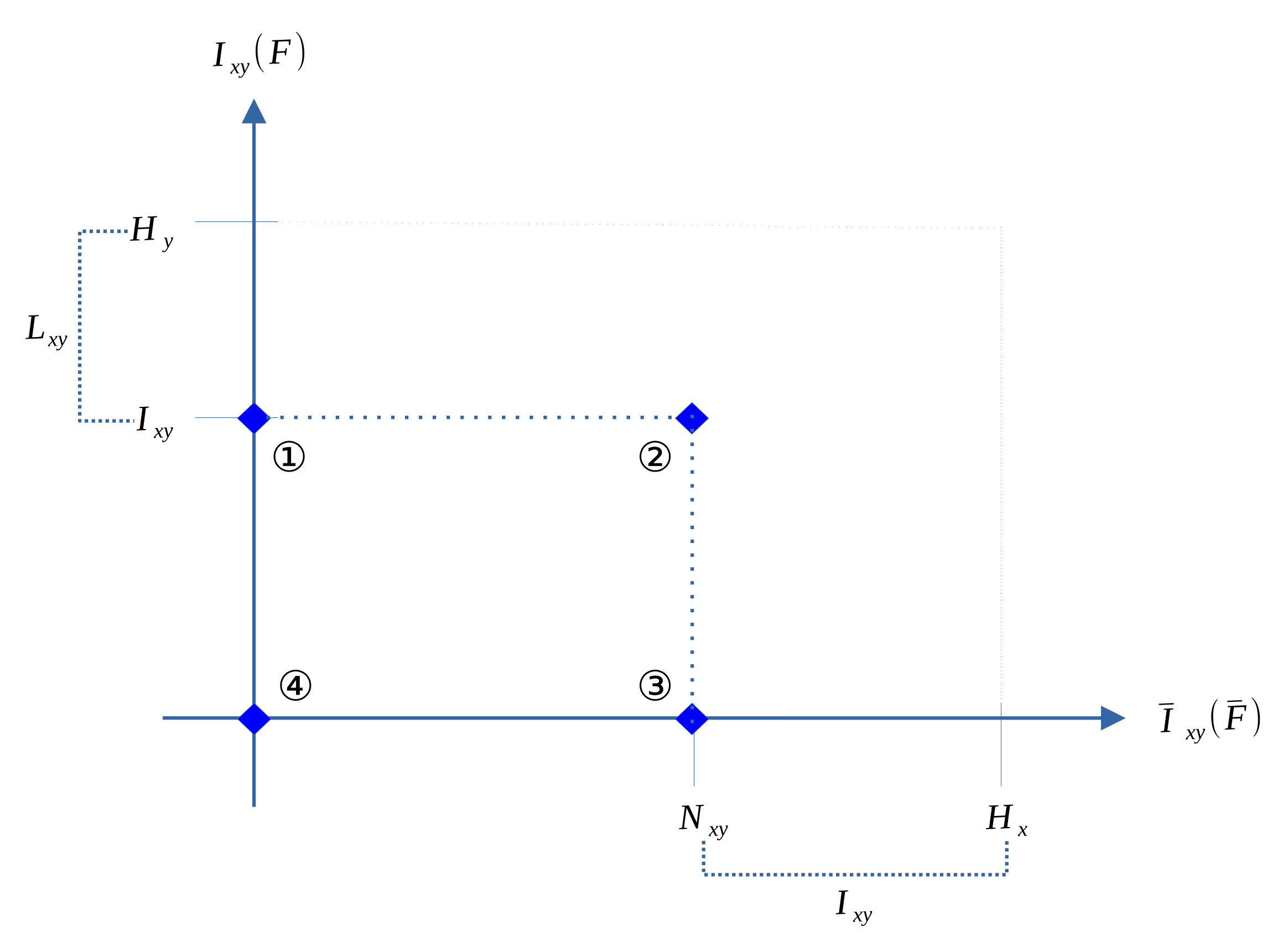}
  \caption{Noise-loss diagram reporting in a Cartesian space the main behavioural patterns found while analysing \IM[\f].} \label{fig:im:deterministic}
\end{figure}
Figure~\ref{fig:im:deterministic} graphically illustrates their positions, which have been reported under the assumption of a deterministic scenario. It is straightforward to find out which coordinates hold for each pattern:

\subsection{Isometrics of \handlemath{\mutualxx[\f]}}

As most of the previous research is centred on the Information Plane (see in particular, \cite{bib:tishby:1999} and \cite{bib:tishby:2015} on this matter), it is useful to draw a bridge between those research findings and noise-loss diagrams. The best way to do that is showing the isometrics of \mutualxx[\f] in the noise-loss space. To this end, one may start from Eq.~\ref{eq:im:deterministic:fundamental-equation} and proceed as follows:
\begin{flalign} \label{eq:im:deterministic:ixx-isometrics}
   \begin{split}
       & \mutualxy[\f] + \noisexy[\f] = \mutualxx[\f] \\
       & \mutualxy[\f] + \noisexy[\f] = \mutualxx[\f] + \noisexy - \noisexy \\
       & \mutualxy[\f] = \big( \noisexy - \noisexy[\f] \big) + \mutualxx[\f] - \noisexy \\       
       & \mutualxy[\f] = \nmutualxy[\nf] + \mutualxx[\f] - \noisexy \\       
   \end{split}
\end{flalign}
The last item of Eq.~\ref{eq:im:deterministic:ixx-isometrics} clearly highlights that the isometrics of \mutualxx[\f] are in fact diagonal lines at 45 degrees. Note that the equation above allows one to calculate the exact value of \mutualxx[\f] for each pair of coordinates, including those that characterise the behavioural patterns previously analysed. For the sake of brevity, let us show how to calculate the value of \mutualxx[\f] for the pattern regarding maximum discriminative capability, which is characterised by the pair of coordinates \pair{\noisexy,\mutualxy}. Substituting these values in the equation derived for representing the isometrics of \mutualxx[\f] allows to write:
\begin{flalign} \label{eq:im:deterministic:ixx:pattern4}
   \begin{split}
       & \mutualxy[\f] = \nmutualxy[\nf] + \big( \mutualxx[\f] - \noisexy \big) \\
       & \mutualxy = \noisexy + \big( \mutualxx[\f] - \noisexy \big) \\
       & \mutualxx[\f] \equiv \mutualxy
   \end{split}
\end{flalign}
This result is not surprising, as $\nmutualxy[\nf] = \noisexy$ implies that all irrelevant information has been removed, leading to $\noisexy[\f] =0$ and hence to $\mutualxx[\f] = \mutualxy[\f] = \mutualxy$.
\begin{figure}[!ht]
  \centering
  \includegraphics[width=0.90\textwidth]{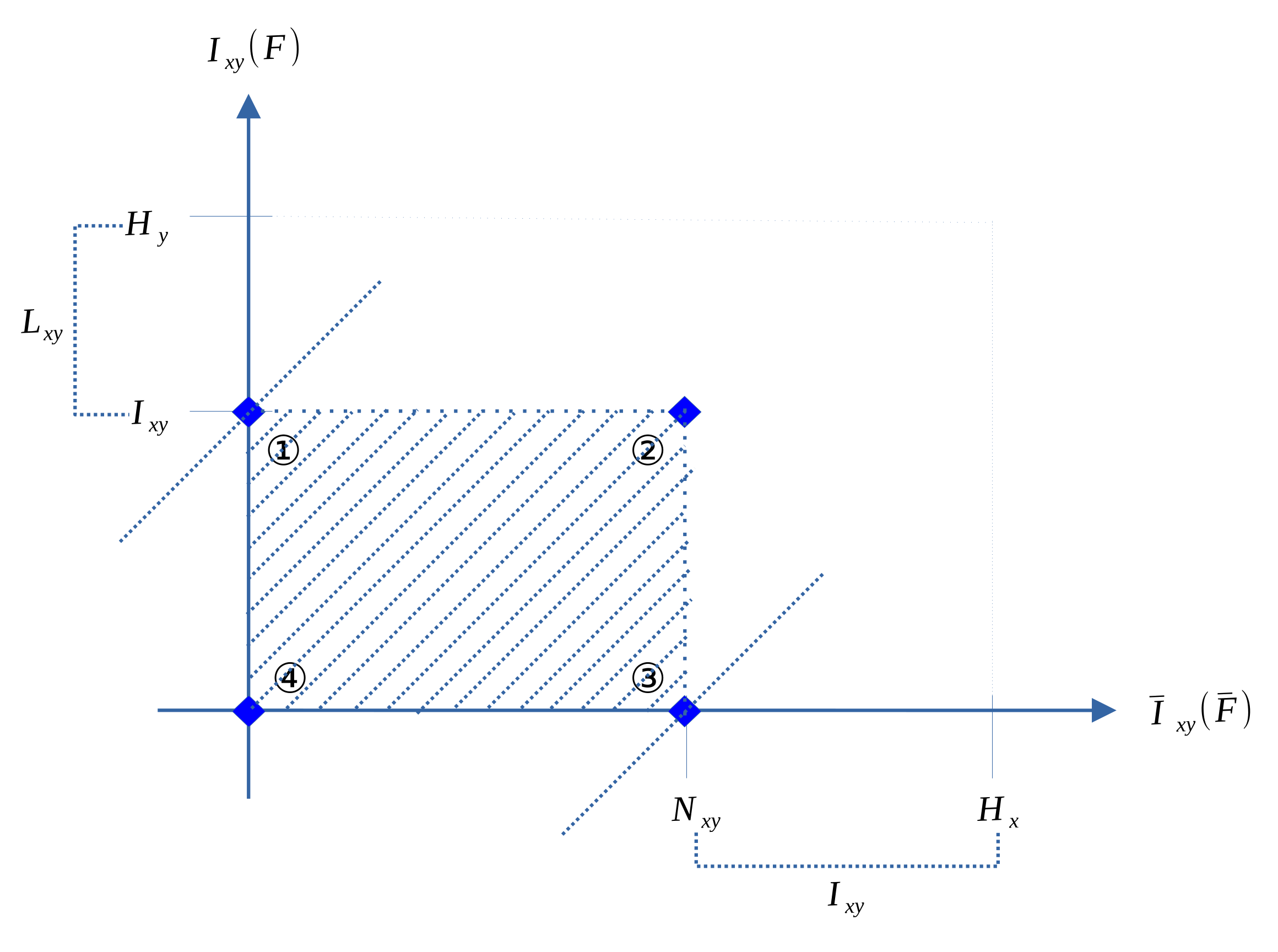}
  \caption{Isometrics of \mutualxx[\f] reported in a noise-loss diagram. The figure highlights that all isometrics are in fact diagonal lines at 45 degrees.} \label{fig:im:deterministic:ixx:isometrics}
\end{figure}

Figure~\ref{fig:im:deterministic:ixx:isometrics} gives a graphical representation of \mutualxx[\f] isometrics, whereas Table~\ref{tab:im:deterministic:ixx:isometrics:summary} reports the values of \mutualxx[\f] for each behavioural pattern.
\begin{table}[ht]
\centering
\centering \small \renewcommand{\arraystretch}{1.5}
\begin{tabular}{l c c c}
 Pattern & \nmutualxy[\nf] & \mutualxy[\f] & \mutualxx[\f] \\
\hline
\phantom{12} \cone \phantom{1} Lossless encoding / compression & 0 & \mutualxy & \Hx \\
\phantom{12} \ctwo \phantom{1} Maximum discriminative capability & \noisexy & \mutualxy & \mutualxy \\
\phantom{12} \cthree \phantom{1} Dummy prediction & \noisexy & 0 & 0 \\
\phantom{12} \cfour \phantom{1} Random prediction & 0 & 0 & \noisexy \\
\hline \\
\end{tabular}
\caption{The value of \mutualxx[\f] for each of the four behavioural patterns previously analysed is reported, together with the corresponding coordinates.} \label{tab:im:deterministic:ixx:isometrics:summary}
\end{table}


\section{Information matrix for non-deterministic scenarios} \label{sec:information-matrix:stochastic}

Non-deterministic scenarios trigger several issues, the most prominent one being whether the overall amount of information may change depending on the noise injected by \f. After clearing this apparent issue, the focus of this section will be on the distinctive features of non-deterministic scenarios. To facilitate the reading, the corresponding analysis will intentionally mimic the one performed for deterministic scenarios.

\subsection{Entropy of the overall information in non-deterministic scenarios}  \label{sub:discussion:stochastic:overall-information}

In principle, switching from a deterministic to a non-deterministic scenario may change the overall information. For instance, thinking about \f as a deterministic transformation \g with additive noise, say \Nx, the following constraints hold on \Hx[\f]:
\begin{equation} \label{eq:im:overall-info:stochastic:constraints}
    \Hx \le max \; \Hx[\f] \le \Hx + \Nx
\end{equation}
Although Eq.~\ref{eq:im:overall-info:stochastic:constraints}  states that one generally cannot predict the effect of the injected noise, the overall information is nevertheless guaranteed to be preserved when \emph{any} of the following conditions is met:
\begin{itemize}
    \item [i)] \Nx is functionally redundant. In particular, \Nx does not increase the overall entropy of \X when thresholding, binning, clipping, or hashing occur on the source (see, for example, \cite{bib:paninski:2003} and \cite{bib:ross:2014} on this matter). 
\item [ii)] \f is a lossy function. This implies that distinct $\langle \G[x], \Nx \rangle$
 pairs can be encoded using the same value.
\item [iii)] \Nx is conditionally unobservable. Meaning that the noise only affects ``uninformative'' parts of \X.
\end{itemize}
Note that the first condition holds because entropy and mutual information estimates require binning and thresholding. The second condition is also satisfied, as the encoding typically maps different input pairs to the same value. The third condition is trivially false, the encoding being a pervasive process that uniformly degrades all components of the source.
However, this disagreement does not affect the final assessment, since satisfying any of the above conditions is sufficient. Consequently, the assumption $max \; \Hx[\f] \equiv \Hx$ is fully justified.

\subsection{Relation between \handlemath{\mutualxy[\f]} and \handlemath{\mutualxx[\f]}}  \label{sub:im:stochastic:ixyf-vs-ixxf}

Under the hypothesis of non-determinism, the relationship between \mutualxx[\f] and \mutualxy[\f] is in full accordance with Eq.~\ref{eq:im:stochastic:fundamental-equation} --repeated here for clarity:
\begin{equation*}
    \mutualxy[\f] + \noisexy[\f] = \Hx[\f] = \mutualxx[\f] + \noisexx[\f]
\end{equation*}
This equation, which lies behind any non-deterministic scenario, highlights that not all the information filtered in is available for prediction, due to the stochastic behaviour of \f. An important consequence is that now \mutualxx[\f] is strictly lower than \Hx[\f], as the equality holds for deterministic scenarios only. 

\down[0.2] Note that Eq.~\ref{eq:im:stochastic:fundamental-equation} has a major importance for three reasons: a)~it allows to move back and forth between backward and forward perspectives, b)~it embeds an important constraint about \noisexy[\f] and \noisexx[\f], and c)~it establishes an equally important constraint between \mutualxy[\f] and \mutualxx[\f].
Let us briefly comment these aspects separately.
\begin{figure}[!ht]
  \centering
  \includegraphics[width=0.65\textwidth]{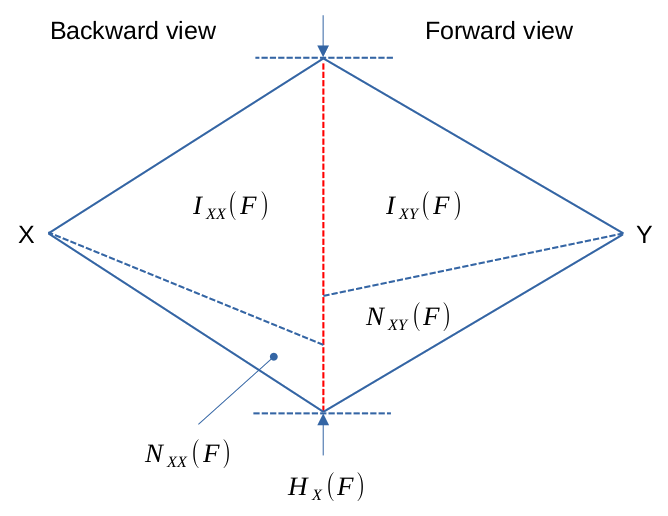}
  \caption{How the information filtered in by \f is split depending on the view (i.e. backward or forward). In particular, the inequalities $\mutualxy[\f] \le \mutualxx[\f]$ and $\noisexx[\f] \le \noisexy[\f]$ are graphically made explicit.} \label{fig:hfx:information-split}
\end{figure}

\down[0.2] \emph{Moving back and forth between perspectives} -- Eq.~\ref{eq:im:stochastic:fundamental-equation} establishes a bridge between backward and forward perspectives. In fact, it is straightforward to express \mutualxy[\f] in terms of \mutualxx[\f] or vice versa. Note that the equation can also be rewritten as $\mutualxx[\f] - \mutualxy[\f] = \noisexy[\f] - \noisexx[\f]$, which highlights the relationships that in this case holds between mutual information and noise.

\down[0.2] \emph{Constraint on \noisexy[\f] and \noisexx[\f]} -- Being independent of \Y, the noise injected by \f is part of the noise that accounts for the irrelevant information left unremoved by \f, so that the constraint $\noisexy[\f] \ge \noisexx[\f]$ holds. In other words, \noisexx[\f] is the minimum value of \noisexy[\f].

\down[0.2] \emph{Constraint on \mutualxy[\f] and \mutualxx[\f]} -- The previous inequality implies that the relevant information retained by \f cannot be greater than the one made available, namely $\mutualxy[\f] \le \mutualxx[\f]$. This assumption is also in accordance with the the \DPI, the intuition being that \FX cannot reveal more about \Y than \X does.

\down[0.2] A graphical representation of the constraints that hold among \mutualxx[\f], \mutualxy[\f], \noisexx[\f] and \noisexy[\f] is also reported in Figure~\ref{fig:hfx:information-split}. 

\subsection{Admissible ranges of noise and loss for non-deterministic scenarios}  \label{sub:im:stochastic:ranges}

Before delving into details, let us recall that non-deterministic assumptions are made for dealing with the uncertainty caused by binning and thresholding, which operate universally across relevant and irrelevant information. 

Under the hypothesis of non-determinism, and assuming that the amount of noise injected by \f is low with respect to the signal embedded in the source \X, the following constraints hold on \noisexy[\f] and \lossxy[\f]:

\begin{align}
    \begin{split}
        \noisexx[\f] &\le \, \noisexy[\f] \le \noisexy \\
        \lossxy + \noisexx[\f] &\le \lossxy[\f] \; \le \Hy
    \end{split}
\end{align}
Note that maximum values remain unchanged with respect to deterministic scenarios, whereas minimum values experience a change. To identify lower bounds for \noisexy[\f] and \lossxy[\f], let us structure the topic according to the hypothesis of additive noise. Specifically, with \g deterministic function and \Nx additive noise, \f can be modelled as follows:
\begin{equation} \label{eq:f-with-additive-noise}
  \FX = \GX + \Nx
\end{equation}

\subsubsection{Lower bound for \noisexy[\f]}

The lower bound of \noisexy[\f] can be ascertained by first recalling that, with \W, \Y random variables and \Z independent noise, the following inequality holds:
\begin{equation} \label{eq:im:noisexy:lower-bound}
     max \; \Big\{ \HH[\W|\Y], \HH[\Z] \Big\} \le \HH[\W+\Z|\Y] \le \HH[\W|\Y] + \HH[\Z]
\end{equation}

Naturally, Eq.~\ref{eq:im:noisexy:lower-bound} can be customised by setting $\W = \GX$ and $\Z = \Nx$. Using the conventional notation for \cHH[\GX|\Y] and \cHH[\FX|\Y], we can write:
\begin{align} \label{eq:im:noisexy:lower-bound:custom}
     max \; \Big\{ \noisexy[\g], \HH[\Nx] \Big\} & \le \noisexy[\f] \le \noisexy[\g] + \HH[\Nx]
\end{align}
The lower bound for \noisexy[\f] can now be easily calculated by imposing the constraint $min \; \noisexy[\g] \equiv 0$ (which depends on the assumption that \g is deterministic) and by considering that \HH[\Nx] is in fact quantified as \noisexx[\f]. In symbols:
\begin{align} \label{eq:im:noisexy:lower-bound:custom:final}
   \begin{split}
     max \; \Big\{ min \; \noisexy[\g], \HH[\Nx] \Big\} & \le min \; \noisexy[\f] \le min \; \noisexy[\g] + \HH[\Nx] \\
     max \; \Big\{ 0, \HH[\Nx] \Big\} & \le min \; \noisexy[\f] \le 0 + \HH[\Nx] \\
     \HH[\Nx] & \le min \; \noisexy[\f] \le \HH[\Nx] \\
     min \; \noisexy[\f] &\equiv \HH[\Nx] \equiv \noisexx[\f]
   \end{split}
\end{align}

\subsubsection{Lower bound for \lossxy[\f]}

As for the minimum of \lossxy[\f], in principle one can only state its range, since the actual value depends on the interaction between the noise injected by \f and the ``signal'' deemed useful to predict \Y. Hence, in general a loss that stands in between \lossxy and $\lossxy + \noisexx[\f]$ is expected. 
Nevertheless, one should not lose sight of the important fact that in \MLPs the assumption of non-determinism is related to the need of implementing binning and thresholding. As a consequence, the information loss caused by the stochasticity of \f is \emph{pervasive}, meaning that it affects all information filtered in by \f. %
Specifically, this implies that \Nx systematically degrades all information in \GX relevant to \Y. Hence, the following inequality holds:
\begin{equation} \label{eq:fxy-with-additive-noise}
           \lossxy[\f] = \HH[\Y|\FX] = \HH[\Y | \GX + \Nx] \ge \lossxy[\g] + \HH[\Nx]
\end{equation}
Eq.~\ref{eq:fxy-with-additive-noise} relies on the key assumption that signal degradation in \f occurs uniformly. In fact, pervasive noise ensures that $min \; \lossxy[\f]$ increases monotonically with \Nx.
This assertion is also in accordance with rate-distortion theory, in which pervasive noise due to uniform quantisation is expected to introduce a fixed entropy penalty per dimension (see for example \cite{bib:cover:2006}).
Despite the fact that \lossxy[\g] is unknown, a non-trivial lower bound for \lossxy[\f] can still be determined, since $min \; \lossxy[\g] = \lossxy$ according to the hypothesis that \g is deterministic. Hence, also recalling that $\HH[\Nx] \equiv \noisexx[\f]$, the following inequality holds:
\begin{equation} \label{eq:fxy-with-additive-noise:lower-bound}
    min \; \lossxy[\f] \ge min \; \lossxy[\g] + \HH[\Nx] = \lossxy + \HH[\Nx] = \lossxy + \noisexx[\f]
\end{equation}

\subsection{Behavioural patterns for non-deterministic scenarios}  \label{sub:im:stochastic:patterns}

The underlying assumption being that the noise injected by \f is low, the behavioural patterns that have been analysed for deterministic scenarios undergo minor changes; which however should be pointed out. As common properties, let us recall that under the hypothesis of non-determinism:
\begin{align} \label{eq:im:stochastic:constraints}
  \begin{split}
    \noisexx[\f] \, & \; \le \; \noisexy[\f] \; \le \; \noisexy \\
    \lossxy + \gammaxx[\f] \, & \; \le \; \lossxy[\f] \; \le \; \Hy\\
  \end{split}
\end{align}

\subsubsection{Pattern \cone (lossless encoding / compression)}

In a non-deterministic scenario, we can only speak about near-lossless encoding / compression, as the minimum value of \lossxy[\f] is greater that \lossxy.~To show that this pattern is compliant with the one regarding the general case, let us derive it by imposing $\noisexy[\f] \equiv \noisexy$ and $\lossxy[\f] = \lossxy + \gammaxx[\f]$:
\begin{align*}
      \nmutualxy[\nf] &= \noisexy - \noisexy[\f] = \noisexy - \noisexy = 0\\
     \nmutualxy[\f] &= \noisexy[\f] = \noisexy \\
     \mutualxy[\nf] &= \dlossxy[\f] = \lossxy + \gammaxx[\f] - \lossxy= \gammaxx[\f]  \\
    \mutualxy[\f] &= \mutualxy - \dlossxy[\f] = \mutualxy - \gammaxx[\f]
\end{align*}
\begin{table}[!ht]
\centering \small \renewcommand{\arraystretch}{1.5}
\begin{tabular}{ l c c }
           & Filtered out by {\f} & Filtered in by {\f} \\
 \cline{2-3}
  Not relevant for \Y & $0$ & $\noisexy$\\ 
  Relevant for \Y & $\gammaxx[\f]$ &  $\mutualxy - \gammaxx[\f]$  \\
  \cline{2-3} \\
\end{tabular}
\caption{How the information embedded in an input source \X is fragmented under the assumption of near-lossless encoding/compression. In this case the first column of \IM[\f] is almost empty, whereas the second one contains all irrelevant information and most of the relevant one.} \label{tab:im:lossless}
\normalsize
\end{table}

In summary, the assumption that \f enacts a near-lossless encoding/compression preserves almost all information embedded in \X, except for the relevant information corrupted by the noise injected by \f. For this reason, the minimum value of \lossxy[\f] in a non-deterministic scenario is $\lossxy + \noisexx[\f]$.

\subsubsection{Pattern \ctwo (maximum discriminative capability)}

When \f is able to provide support for achieving the highest level of discriminative capability, almost all relevant information is retained, whereas almost all irrelevant information is removed. Looking back at the ranges that hold for \noisexy[\f] and \lossxy[\f] in case of stochastic transformations, the pattern reported in Table~\ref{tab:im:stochastic:max-discriminative-ability} holds.
\begin{table}[!ht]
\centering \small \renewcommand{\arraystretch}{1.5}
\begin{tabular}{ l c c }
           & Filtered out by {\f} & Filtered in by {\f} \\
 \cline{2-3}
  Not relevant for \Y & $\noisexy - \noisexx[\f]$ & \noisexx[\f]\\ 
  Relevant for \Y & \gammaxx[\f] &  $\mutualxy - \gammaxx[\f]$  \\
  \cline{2-3} \\
\end{tabular}
\caption{How the information embedded in an input source \X is fragmented in the case in which \f reaches the highest performance in a non-deterministic scenario. Note that in this case the elements of the secondary diagonal are not null, due to the corruption enacted by the stochasticity of \f.} \label{tab:im:stochastic:max-discriminative-ability}
\normalsize
\end{table}
To show that this pattern is compliant with the one of the general case, let us impose the constraints $\noisexy[\f] = \noisexx[\f]$ and $\lossxy[\f] = \lossxy + \gammaxx[\f]$:
\begin{align*}
     \nmutualxy[\nf] &= \noisexy - \noisexy[\f] = \noisexy - \noisexx[\f] \\
     \mutualxy[\nf] &= \noisexy[\f] = \noisexx[\f] \\
     \nmutualxy[\f] &= \dlossxy[\f] = \lossxy + \gammaxx[\f] - \lossxy = \gammaxx[\f] \\
     \mutualxy[\f] &= \mutualxy - \dlossxy[\f] = \mutualxy - \gammaxx[\f] \\
\end{align*}
Naturally, in the event that \X contains all information required to predict \Y, Table~\ref{tab:im:stochastic:max-discriminative-ability} should be rewritten by substituting \lossxy with $0$, thus giving rise to a near oracle-like behaviour.

In summary, in a non-deterministic scenario the pattern of maximum discriminative capability can only be approximated. In fact, due to the corruption enacted by the stochasticity of \f,  some irrelevant information is left unremoved, whereas some relevant information is lost. 
In particular, with respect to the deterministic scenario, the noise left unremoved is \noisexx[\f], whereas the amount of relevant information filtered in decreases to $\mutualxy - \noisexx[\f]$. As for the oracle-like behaviour, it can only be approximated due to the stochasticity of \f.

\subsubsection{Pattern \cthree (dummy prediction)}

In principle, this pattern assumes all information be filtered out by \f. However, as reported in Table~\ref{tab:im:stochastic:relevant-info-lost}, the unavoidable noise injected by \f enacts a minor difference, a small part of irrelevant information being retained.
\begin{table}[!ht]
\centering \small \renewcommand{\arraystretch}{1.5}
\begin{tabular}{ l c c }
           & Filtered out by {\f} & Filtered in by {\f} \\
 \cline{2-3}
  Not relevant for \Y & $\noisexy - \noisexx[\f]$ & $\noisexx[\f]$\\ 
  Relevant for \Y & $\mutualxy$ &  $0$  \\
  \cline{2-3} \\
\end{tabular}
\caption{How the information embedded in an input source \X is fragmented in case of complete loss of relevant information. As expected, the noise injected by \f is left unremoved.} \label{tab:im:stochastic:relevant-info-lost}
\normalsize
\end{table}
To show that this pattern is compliant with the one regarding the general case, let us derive it by imposing $\noisexy[\f] = \noisexx[\f]$ and that $\lossxy[\f] = \Hy$:
\begin{align*}
      \nmutualxy[\nf] &= \noisexy - \noisexy[\f]= \noisexy - \noisexx[\f] \\
     \nmutualxy[\f] &= \noisexy[\f] = \noisexx[\f] \\
      \mutualxy[\nf] &= \dlossxy[\f] = \mutualxy \\
     \mutualxy[\f] &= \mutualxy - \dlossxy[\f] = \mutualxy - \mutualxy = 0 \\
\end{align*}

In summary, this pattern is characterised by the inability to retain relevant information while removing the irrelevant one (except for the unavoidable noise injected by \f).

\subsubsection{Pattern \cfour (random prediction)}

This pattern assumes that all relevant information is removed by \f, whereas all irrelevant information is left unremoved. 
The corresponding \IM[\f] is identical to the one already analysed in the deterministic scenario.

\subsection{Isometrics of \handlemath{\mutualxx[\f]}}

The isometrics of \mutualxx[\f] in a noise-loss diagram under the hypothesis of non-determinism can be made explicit starting from Eq.~\ref{eq:im:stochastic:fundamental-equation} and proceeding as follows:
\begin{flalign} \label{eq:im:stochastic:ixx:isometrics}
   \begin{split}
       & \mutualxy[\f] + \noisexy[\f] = \mutualxx[\f] + \noisexx[\f] \phantom{\Big( \Big)}\\
       & \mutualxy[\f] + \noisexy[\f]= \mutualxx[\f] + \noisexx[\f]  + \noisexy - \noisexy \\
       & \mutualxy[\f] = \Big( \noisexy - \noisexy[\f] \Big) + \mutualxx[\f] + \noisexx[\f] - \noisexy \\       
       & \mutualxy[\f] = \nmutualxy[\nf] + \mutualxx[\f] + \noisexx[\f] - \noisexy \\       
   \end{split}
\end{flalign}
The last item of Eq.~\ref{eq:im:stochastic:ixx:isometrics} clearly highlights that the isometrics of \mutualxx[\f], in a noise-loss diagram, are still diagonal lines at 45 degrees. As made under the hypothesis of deterministic \f, the equation above allows to calculate the exact value of \mutualxx[\f] for each pair \pair{\nmutualxy[\nf],\mutualxy[\f]}, including those that characterize the specific behavioural patterns previously analysed.
Let us only calculate the value of \mutualxx[\f] for the pattern regarding maximum discriminative capability, which is characterised by the pair of coordinates \pair{$\,\noisexy - \noisexx[\f]$,$\,\mutualxy-\gammaxx[\f]$}. Substituting these values in the equation derived for representing the isometrics of \mutualxx[\f], one obtains:
\begin{flalign} \label{eq:im:stochastic:ixx:max-performance}
   \begin{split}
       & \mutualxy[\f] = \nmutualxy[\nf] + \mutualxx[\f] + \noisexx[\f] - \noisexy \\
       & \mutualxy - \gammaxx[\f] = \noisexy - \noisexx[\f] + \mutualxx[\f] + \noisexx[\f] - \noisexy \\
       & \mutualxy - \gammaxx[\f] = \mutualxx[\f] \\
       & \mutualxx[\f] = \mutualxy - \gammaxx[\f]
   \end{split}
\end{flalign}
The result is still not surprising. Since in this case $\noisexy[\f] = \noisexx[\f]$, this implies (from Eq.~\ref{eq:im:stochastic:fundamental-equation}) that $\mutualxx[\f] = \mutualxy[\f] = \mutualxy - \gammaxx[\f]$.
\begin{figure}[!ht]
  \centering
  \includegraphics[width=0.90\textwidth]{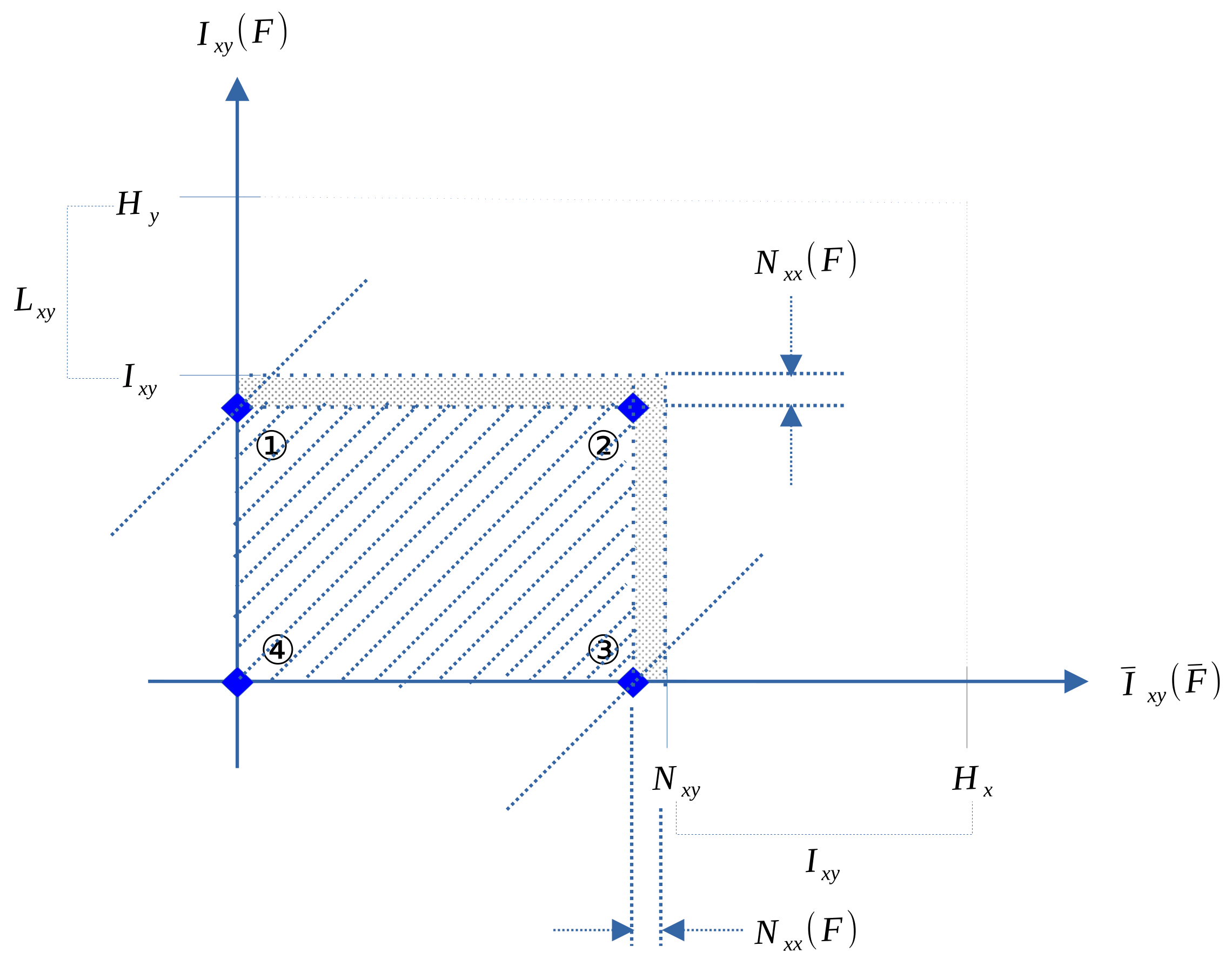}
  \caption{Isometrics of \mutualxx[\f] reported in a noise-loss diagram. The figure highlights that all isometrics are still diagonal lines at 45 degrees, whose origin depends only on the amount of irrelevant information and on the noise injected by \f.} \label{fig:im:stochastic:ixx:isometrics}
\end{figure}
For the sake of completeness, Figure~\ref{fig:im:stochastic:ixx:isometrics} gives a graphical representation of \mutualxx[\f] isometrics, whereas Table~\ref{tab:im:stochastic:ixx:isometrics:summary} reports the values of \mutualxx[\f] for each behavioural pattern deployed in a non-deterministic scenario.  
\begin{table}[!ht]
\centering
\centering \small \renewcommand{\arraystretch}{1.5}
\setlength{\tabcolsep}{6pt}
\begin{tabular}{l c c c}
 Pattern & \nmutualxy[\nf] & \mutualxy[\f] & \mutualxx[\f] \\
\hline
\cone \phantom{:} Lossless enc./compr. & 0 & \mutualxy - \gammaxx[\f] & $\Hx - 2 \cdot  \gammaxx[\f]$ \\
\ctwo \phantom{:} Max discr. capability & \noisexy - \noisexx[\f] & \mutualxy - \gammaxx[\f] & \mutualxy - \gammaxx[\f] \\
\cthree \phantom{:} Dummy prediction & \noisexy - \noisexx[\f] & 0 & 0 \\
\cfour \phantom{:} Random prediction & 0 & 0 & \noisexy - \noisexx[\f] \\
\hline \\
\end{tabular}
\caption{The table reports the value of \mutualxx[\f] corresponding to each of the four behavioural patterns previously analysed. For the sake of completeness, also the corresponding coordinates of the noise-loss diagram are reported.} \label{tab:im:stochastic:ixx:isometrics:summary}
\end{table}


\section{Optimisation strategies} \label{sec:optimisation-strategies}

The analysis previously made on specific behavioural patterns is not an end in itself. In fact, it can be highly helpful in the task of building a model with optimal or suboptimal performance. This section is focused on this matter. In particular, first some straightforward optimisation strategies are derived from \IM[\f]; then, a general parametric optimisation strategy is devised and commented. 

\subsection{Straightforward optimisation strategies}

Using \IM[\f] as reference, optimisation strategies may be focused on: i)~limiting the loss of relevant information, ii)~promoting the removal of irrelevant information, or iii)~promoting the removal of irrelevant information while limiting the loss of the relevant one.

It is worth pointing out in advance that hereinafter terms not subject to optimisation will be highlighted with the tag ``NSO'' and removed.

\down[0.2] \emph{Focus on limiting the loss of relevant information}.
This strategy can be achieved by maximising \mutualxy[\f] or, equivalently, by minimising \mutualxy[\nf]. Selecting the latter, the following equation holds:
\begin{align} \label{eq:optimisation:focus:relevant-info}
  \begin{split}
  \Fstar & = \argmin_{\f} \Big( \; \mutualxy[\nf] \; \Big) \\
              & = \argmin_{\f} \Big( \; \lossxy[\f] - \nso{\lossxy} \; \Big) \\
              & = \argmin_{\f} \Big( \; \lossxy[\f] \; \Big)
  \end{split}
\end{align}

Note that both patterns \cone and \ctwo are aimed at limiting the loss of relevant information. However, the outcomes of this strategy would likely be dominated by pattern \cone, due to the lack of control on noise removal.

\down[0.2] \emph{Focus on removing irrelevant information}.
This strategy can be achieved by maximising \nmutualxy[\nf] or, equivalently, by minimising \nmutualxy[\f]. Selecting the latter, the following equation holds:
\begin{align}  \label{eq:optimisation:focus:irrelevant-info}
  \begin{split}
\Fstar & = \argmin_{\f} \Big( \; \nmutualxy[\f] \; \Big) \\
            &= \argmin_{\f} \Big( \; \noisexy[\f] \; \Big)
  \end{split}
\end{align}

Note that both patterns \ctwo and \cthree are aimed at removing irrelevant information. However, the outcomes of this strategy would likely be dominated by pattern \cthree, due to the lack of control on loss avoidance.

\down[0.2] \emph{Focus on removing irrelevant information while limiting the loss of the relevant one}.
Being a combination of the previous ones, this strategy can be achieved by maximising the primary diagonal or, equivalently, by minimising the secondary one. Selecting the latter, the following equation holds:
\begin{align} \label{eq:optimisation:focus:both}
  \begin{split}
  \Fstar &= \argmin_{\f} \Big( \; \nmutualxy[\f] + \mutualxy[\nf] \; \Big) \\
              &= \argmin_{\f} \Big( \; \noisexy[\f] + \big( \, \lossxy[\f] - \nso{\lossxy} \, \big) \; \Big) \\
              &= \argmin_{\f} \Big( \; \noisexy[\f] + \lossxy[\f] \; \Big) \\
  \end{split}
\end{align}

Note that only pattern \ctwo takes care of removing irrelevant information while limiting the loss of the relevant one.

\subsection{Parametric optimisation strategy}

The simplest generalisation that encompasses all the above strategies can be obtained by introducing a hyperparameter, say $\alpha \in[0,1]$, devised to control to what extent the focus should be set on removing irrelevant information or on retaining the relevant one. Taking the secondary diagonal of \IM[\f] as reference, a parametric optimisation strategy can be defined as follows:
\begin{align} \label{eq:optimisation:parametric}
\begin{split}
  \Fstar &= \argmin_{\f} \Big(  \; \alpha \cdot \noisexy[\f] + ( 1 - \alpha) \cdot \big( \, \lossxy[\f] - \nso{\lossxy} \, \big) \; \Big) \\
               &= \argmin_{\f} \Big(  \; \alpha \cdot \noisexy[\f] + ( 1 - \alpha) \cdot \lossxy[\f] \; \Big)\\
\end{split}
\end{align}
Let us analyse with detail the scenarios that occur with varying $\alpha$:

\begin{itemize}
\item[--]  $\alpha = 0$. The focus is on retaining relevant information (i.e. on limiting the loss), whereas no constraints hold aimed at removing the irrelevant one (see also Eq.~\ref{eq:optimisation:focus:relevant-info}). Any algorithm guided by this heuristic may find a minimum along the straight line that links pattern \cone with pattern \ctwo. However, this optimisation strategy would most likely lead to pattern \cone, thereby promoting the generation of predictors that implement rote learning, namely of a predictor unable to perform generalisation.

\item[--] $0 < \alpha < 0.5$. In this range the aim of keeping relevant information prevails on the need of removing the irrelevant one. The more $\alpha$ is close to $0$, the more this heuristic is enforced.
\item[--]  $\alpha = 0.5$. The focus is equally set on retaining relevant information and concurrently removing the irrelevant one. This strategy is aimed at reaching the maximum discriminative capability by controlling noise and loss with the same care (see pattern \ctwo and also Eq.~\ref{eq:optimisation:focus:both}).
\item[--]   $0.5 < \alpha < 1$.  In this range the aim of removing irrelevant information prevails on the need of retaining the relevant one. The more $\alpha$ is close to $1$, the more this heuristic is enforced.
\item[--] $\alpha = 1$. The focus is on removing irrelevant information, whereas no constraints hold aimed at retaining the relevant one (see also Eq.~\ref{eq:optimisation:focus:irrelevant-info}). Any algorithm guided by this heuristic may find a minimum along the straight line that links pattern \ctwo with pattern \cthree. However, the most likely outcome would be coincident with pattern \cthree, which promotes the finding of dummy predictors, namely of predictors that emit a constant regardless of the actual input.

\end{itemize}

It is noteworthy that the analysis above omits any discussion of pattern \cfour. This should not be surprising, since attempting to maximize the secondary diagonal lacks any theoretical justification.

\subsection{Similarities with the optimisation strategy suggested in the \IB framework}

The reference optimisation strategy devised for the \IB framework (see in particular~\cite{bib:tishby:2015} on this matter) consists of minimising the following Lagrangian --where \hlayer represents the dividing line between encoding and decoding:
\begin{equation} \label{eq:tishby:lagrangian}
  \mathcal{L}_{IB} = \mutual{\X;\hlayer} - \beta \cdot \mutual{\hlayer;\Y}
\end{equation}

One may wonder about the apparent difference between the optimisation strategy relating to the \IB framework (Eq.~\ref{eq:tishby:lagrangian}) and the one proposed in this article (Eq.~\ref{eq:optimisation:parametric}). A schematic comparison follows hereinafter. 

\down[0.2] With $\hlayer \equiv \FX$, the goal is to maximise the performance of the predictor, with varying \f (note that the equivalence does not restrict generality, as Section~\ref{sec:information-flow-in-mlps} will show that any layer of an \MLP can be taken as reference). 
As first step,  starting from Eq.~\ref{eq:im:stochastic:fundamental-equation}, let us reformulate \mutualxx[\f] as follows:
\begin{align} \label{eq:mutual-info:as-mutualxx}
\begin{split}
  \mutualxx[\f] &= \mutualxy[\f] + \noisexy[\f] - \noisexx[\f]
\end{split}
\end{align}
Hence, we can write (terms not subject to optimisation are progressively removed):
\begin{align} \label {eq:optimisation:parametric:lagrangian}
   \begin{split}
     \mathcal{L}_{IB} &= \mutualxx[\f] - \beta \cdot \mutualxy[\f] \\
                                     &= \Big( \; \mutualxy[\f] + \noisexy[\f] - \nso{\noisexx[\f]} \; \Big) - \beta \cdot \mutualxy[\f] \, \big) \\
                                    &= \noisexy[\f] +  (1 - \beta) \cdot \mutualxy[\f] \\
                                    &= \noisexy[\f] +  (1 - \beta) \cdot \Big( \; \nso{\Hy} - \lossxy[\f] \; \Big) \\
                                    &= \noisexy[\f] +  (\beta - 1) \cdot \lossxy[\f] \quad \quad \beta \ge 0
   \end{split}
\end{align}
Note that \noisexx[\f] has been removed from the functional form, as the noise injected by \f is not directly optimizable in practice.
For comparison purposes, let us reformulate Eq.~\ref{eq:optimisation:parametric}, with the goal of mimicking the result just found:
\begin{align}  \label {eq:optimisation:parametric:reformulated}
  \begin{split}
  \Fstar & = \argmin_{\f} \Big( \; \alpha \cdot \noisexy[\f] + (1 - \alpha) \cdot \lossxy[\f] \; \Big)  \quad \quad \quad 0 \le \alpha \le 1 \\
              & = \argmin_{\f} \Big( \; \noisexy[\f] + \big( \, \frac{1}{\alpha} - 1 \, \big) \cdot \lossxy[\f] \; \Big) \\
              & = \argmin_{\f} \Big( \; \noisexy[\f] + \big( \, \beta - 1 \, \big) \cdot \lossxy[\f] \; \Big) \quad \quad \quad \quad \beta \eqdef 1/\alpha \ge 1
  \end{split}
\end{align}
Eq.~\ref{eq:optimisation:parametric:lagrangian} and \ref{eq:optimisation:parametric:reformulated} highlight that there is a substantial identity between the two optimisation strategies, except for the constraint that applies to the parameter $\beta$.
However, this is not a major issue, as several studies performed on the \IB framework suggest to use $\beta \ge 1$ to avoid degenerate solutions.
In particular, \authors{Tishby and Zaslavsky}~\cite{bib:tishby:2015} observe that for $\beta < 1$ the penalty on compression is weaker than the reward for discarding information. Moreover, in a 2017 article, \authors{Shwartz-Ziv and Tishby} note that $\beta \ge 1$ is needed to avoid collapsing to the ``no information'' solution.
In fact, perhaps contrary to intuition, the motivation for keeping $\beta \ge 1$ is that otherwise (as pointed out by Eq.~\ref{eq:optimisation:parametric:lagrangian}) the optimisation would start working in the \emph{opposite} direction, trying to maximise the loss of relevant information together with the minimisation of the irrelevant one left unremoved.
Should the need arise, the above findings clearly highlight that \IM[\f] is in fact \emph{the} starting point for any sound proposal concerning optimisation strategies.


\section{Information flow in multilayer perceptrons} \label{sec:information-flow-in-mlps}

This section is focused on using \IM[\f] for analysing the behaviour of an \MLP. 
Before delving into details, a brief comment on the underlying scenario is needed; specifically, whether it is framed according to a deterministic or non-deterministic assumption.
Apart from the inherent approximations related to floating-point computations, in principle any layer of a standard \MLP performs a non-linear, yet deterministic, transformation over the input provided by the previous layer (or by the input layer). Although this concept may be considered trivial, it is not. In fact, most often the studies in this field do not take into account the deterministic nature of the transformation enforced by a standard \MLP, layer by layer, with few exceptions (e.g. the work of \authors{Strouse and Schwab}\cite{bib:strouse:2016}).
On the other hand, as already pointed out, the rationale for assuming a non-deterministic scenario strictly depends on the need of estimating entropy and mutual information.
Hence, switching to non-determinism is a choice that depends on pragmatic considerations (i.e. the construction of a convenient framework for entropy and mutual information estimation) rather than on theoretical necessity.

\down[0.2] Having clarified this, the behaviour of an \MLP is analysed hereinafter assuming that non-determinism holds.

\subsection{Common definitions regarding \MLPs}

\begin{figure}[!ht]
  \centering
  \includegraphics[width=0.90\textwidth]{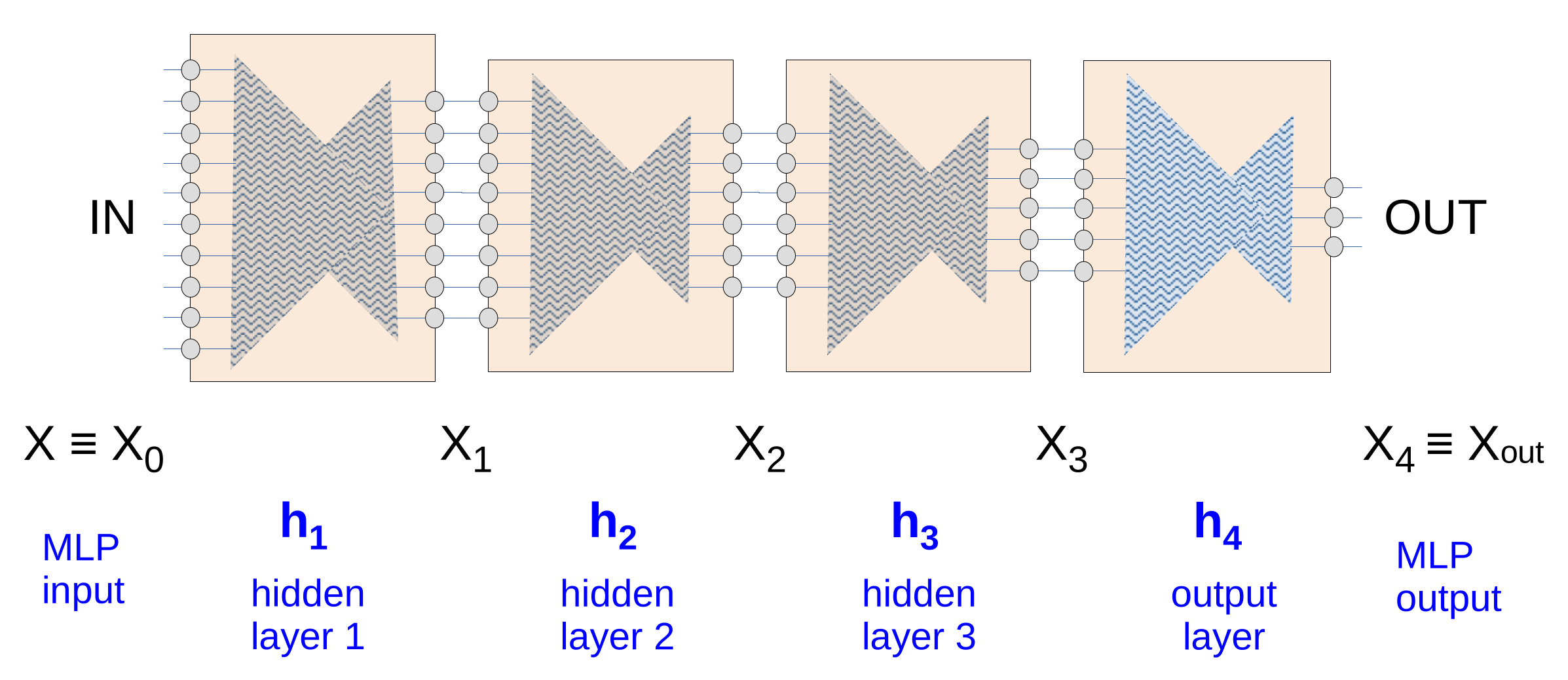}
  \caption{An \MLP with three hidden layers. In particular, the \kth layer is denoted with \layer[k], whereas the corresponding output (from the input up to layer \layer[k]) is denoted with \Xstep[k]. The figure also highlights the difference between the transformation performed by a layer and its output. } \label{fig:mlp:layers}
\end{figure}

Be \mlp a generic \MLP with \numlayers hidden layers, numbered from 1 to \numlayers. The \kth hidden layer of \mlp is denoted with \layer[k]. As a consequence of the adopted numbering scheme, the input and the output layers of \mlp may also be denoted with \layer[0] and \layer[\numlayers+1], respectively. As an example, Figure~\ref{fig:mlp:layers} reports an \MLP with three hidden layers ($\numlayers=3$), highlighting that the output of layer \layer[k] becomes an input for layer \layer[k+1] with varying $k$ from $1$ to $3$.
The figure also differentiates between the transformation performed by a layer and the corresponding output. For instance, the random variable \Xstep[2], which is the output of the hidden layer \layer[2], results from the transformation applied to the input by the pipeline of layers  \layer[1] and \layer[2]. The random variable $\X \equiv \Xstep[0]$ represents the input of \mlp, whereas $\Xstep[4] \equiv \Xstep[out]$ represents its output. \\
The information processing that occurs across layers can be highlighted by performing a step-by-step analysis.
Let us denote with \Xstep[k] the random variable that holds at the \kth layer and with \layer[k] the transformation enforced therein. With these settings in mind, a constructive definition of \Xstep[k] obeys the following rule:
\begin{equation} \label{eq:mlp:information-flow:X:recursive}
   \Xstep[k] =
      \begin{cases} 
         \X & \quad k=0 \\
         \layer[k](\Xstep[k-1]) & \quad k=1,2, ... ,\numlayers+1
      \end{cases}
\end{equation}
As nothing prevents from seeing the transformation that generates \Xstep[k] from \X as an encoder, let us put into practice this insight by defining \f[k]:
\begin{equation} \label{eq:mlp:information-flow:F:unfolded}
     \f[k] =
       \begin{cases}
          \layer[1] & \quad k=1 \\
          \layer[k] \circ \f[k-1] & \quad k=2, ..., \numlayers+1 \\
       \end{cases}
\end{equation}

\down[0.2] In so doing, the definition given in Eq.~\ref{eq:mlp:information-flow:X:recursive} can be unfolded as follows:
\begin{equation} \label{eq:mlp:information-flow:X:unfolded}
     \Xstep[k] =
       \begin{cases}
          \X & \quad k=0 \\
          \Fstep[k]{\X} & \quad k=1,2, ..., \numlayers+1 \\
       \end{cases}
\end{equation}
With the goal of representing the overall transformation \FX as a pipeline of encoding and decoding, a decoder (say \Dname[k]) can be defined as follows: 
\begin{equation} \label{eq:mlp:information-flow:G:unfolded}
     \Dname[k] =
       \begin{cases}
          \layer[m+1] & \quad k=m+1 \\
         \Dname[k+1] \circ  \layer[k] & \quad k=1,2, ..., \numlayers \\
       \end{cases}
\end{equation}
The definition above makes it possible to rewrite the overall transformation \FX as a combination of encoding and decoding, with varying $k$, as follows:
\begin{equation} \label{eq:mlp:information-flow:encoder-decoder}
    \FX = \Dname[k+1] \circ \Fstep[k]{\X} \quad k=1,2, ..., m
\end{equation}

\subsection{Extending the analysis to hidden layers}

The previous subsection pointed out that any specific layer of an \MLP can be seen as an encoder followed by a decoder, regardless of the selected ``cut point''. 
In fact, \emph{any} layer of an \MLP can be used as a sort of pivot, from which one may look backward to the source or forward to the target. 
Hence, we are now able to quit the hypothesis, made in Section~\ref{sec:materials-and-methods}, that \f should be considered a black-box. 
As a consequence, all results obtained with \f can be used, \emph{at no cost}, to investigate any layer of an \MLP. 

The notation \fk will be used hereinafter to denote the transformation enacted up to the $k$-th layer. Naturally, this notation extends to all terms devised during the analysis regarding the overall transformation, the only difference being that \f should be substituted by \f[k].

\subsubsection{Inequalities that hold within and across \MLP layers}

\down[0.2] The \DPI principle can be applied to any stage of the processing chain of an \MLP. In particular, the following inequality holds:
\begin{equation} \label{eq:mutualxx-vs-mutualxy}
    \mutualxy[\fk] \le \mutualxx[\fk] \quad \kstep = 1, 2, ..., \numlayers+1
\end{equation}
Eq.~\ref{eq:mutualxx-vs-mutualxy}  can be useful for investigating the informative content of a generic layer. The right-hand component represents the information available at the \kstep-th layer, whereas the one on the left shows to what extent the output of the layer in question is expected to support the task of predicting the target.

\down[0.2] \mutualxx[\f] and \mutualxy[\f] are also involved in two important processing chains. In symbols:
\begin{align} \label{eq:dpi:mutual-info:across-mlp}
\begin{split}
    \Hx &\ge \mutualxx[{\f[1]}] \ge \mutualxx[{\f[]}] \; \; ... \; \; \ge \mutualxx[{\f[\numlayers]}] \ge \mutualxx[\f] \\
    \mutualxy &\ge \mutualxy[{\f[1]}] \ge \mutualxy[{\f[]}] \; \; ... \; \; \ge \, \mutualxy[{\f[\numlayers]}] \ge \, \mutualxy[\f]
   \end{split}
\end{align}
Eq.~\ref{eq:dpi:mutual-info:across-mlp} highlights that, during the training process, available and relevant information are monotonically decreasing across layers, with Eq.~\ref{eq:mutualxx-vs-mutualxy} concurrently ensuring that the former is always greater than or equal to the latter at any layer.

\down[0.2] Similar considerations allows to characterise noise and loss chains, which obey the following inequalities:
\begin{align} \label{eq:dpi:noise-and-loss:across-mlp}
\begin{split}
    \noisexy &\ge \noisexy[{\f[1]}] \ge \noisexy[{\f[]}] \; \; ... \; \; \ge \noisexy[{\f[\numlayers]}] \ge \noisexy[\f] \\
    \lossxy &\le \lossxy[{\f[1]}] \le \lossxy[{\f[]}] \; \; \; ... \; \;  \le \lossxy[{\f[\numlayers]}] \le \lossxy[\f]
   \end{split}
\end{align}
Eq.~\ref{eq:dpi:noise-and-loss:across-mlp} highlights that, during the training process, the input source is (hopefully) made more and more compatible with the target as the information flows across the layers, at the risk of losing some relevant information along the way.

\subsubsection{Using \handlemath{\IM[\f]} for analysing the information flow}

To highlight the shift from a black-box to an in-layer perspective, the information matrix that holds at the $k$-th layer is denoted by \IM[\fk].

With  $r$ and $s$ obeying the constraint $0 \le r \le s \le \numlayers+1$, the information processing chain reported in Eq.~\ref{eq:dpi:noise-and-loss:across-mlp}, concerning the removal of irrelevant information and the potential loss of relevant one, can be expressed in a compact form by defining the operator ``$\ge$'' for \IM as follows:
\begin{equation} \label{eq:im:operator-greater-than-or-equal}
    \IM[{\f[r]}] \ge \IM[{\f[s]}] \iff \Big( \noisexy[{\f[r]}] \ge \noisexy[{\f[s]}] \Big) \land \Big( \lossxy[{\f[r]}] \le \lossxy[{\f[s]}] \Big)
\end{equation}

According to the given definition, a typical information chain based on \IM would be:
\begin{align} \label{eq:im:info-matrix:chain}
        \IM[\X] & \ge \IM[{\f[1]}] \ge \IM[{\f[2]}] \; \; \; ... \; \; \; \ge \IM[{\f[m]}] \ge \IM[\f]
\end{align} 
As for the notation \IM[\X], it has been chosen to better highlight that any processing activity starts with all information located at the ``filtered-in'' side, as shown hereinafter:
\down[0.2] %
{ \par\centering %
  \small \renewcommand{\arraystretch}{1.5}
\begin{tabular}{ l c c }
           & Filtered out & Filtered in \\
 \cline{2-3}
  Not relevant for \Y & 0 & \noisexy \\ 
  Relevant for \Y & 0 &  \mutualxy \\
  \cline{2-3} \\
\end{tabular} \par}

Note that, due to \DPI, the inequalities reported in Eq.~\ref{eq:im:info-matrix:chain} are trivially true for any \MLP chain. However, analysing a chain content may be very useful for studying to what extent noise removal and loss of relevant information occurred during the training process, layer by layer.
As a final comment on this matter, observe that the chain still points out that the irrelevant information is expected to be progressively removed, with the caution that some relevant information may concurrently be lost across layers.

\subsection{\MLP training and the compression issue}

In \cite{bib:shwartz-ziv:2017}, \authors {Shwartz-ziv and Tishby} claim that fitting followed by compression appears to be a universal mechanism that occurs during the training phase of an \MLP. In fact, there has been extensive debate over this claim. 
An important contribution was made by \authors{Saxe et al.}~\cite{bib:saxe:2018}, who demonstrate that compression may or may not arise, depending on the activation function used. In particular, compression occurs while using the sigmoid as activation function, whereas it is absent with ReLU.
Further research in this field reinforces the view that compression is not a universal mechanism.
In particular, \authors{Amiad and Geiger}~\cite{bib:amjad:2019} claim that compression is due to the noise relating to stochastic gradient descent, rather than to the \IB principle itself. Moreover, \authors{Schiemer and Ye}~\cite{bib:schiemer:2020} argue that compression emerges only with explicit regularisation (e.g., enforcing techniques such as random dropout and weight decay). See also the review article~\cite{bib:piran:2021} on this matter.

In this position article an alternative view is proposed, related to the inherent characteristics of the training process.
Fitting is strictly related to the adopted training strategy, whereas compression is typically enforced at the architectural level. Abstracting away implementation details, the former results from updating \MLP weights in accordance with the adopted training strategy and the latter occurs as the number of neurons typically decreases across layers. Despite the variety of proposals, the most likely hypothesis is that these mechanisms occurs jointly rather than one following the other. The motivation is clear: at each step both fitting and compression act concurrently, as one is complementing the other.
This inspiring principle is even clearer when moving beyond standard backpropagation to embrace other approaches.
In fact, updating weights across layers of a fixed \MLP architecture is not the only training strategy that can be put into practice. In particular, layerwise training (see for instance \cite{bib:bengio:2007}, \cite{bib:gomez:2017}, and \cite{bib:armano:2020}) performs training layer by layer instead of repeatedly working on the whole \MLP architecture. 
This hands-on perspective sheds light on the overall transformation enforced by an \MLP, suggesting that a chain aimed at \emph{adapting} the input source to the target is actually put into practice.


\section{Conclusions and future work} \label{sec:conclusions}

In this article an in-depth analysis has been performed on the way multilayer perceptrons process the input source in supervised settings.
The analysis is centred on the concept of information matrix, which has been devised and then used as formal tool for understanding the aetiology of optimisation strategies and for studying how the information flows across layers.
It has also been pointed out that a multilayer perceptron functions as an adaptor, progressively transforming the input source across its layers to mimic the target.
As for future work, several experiments are in progress, aimed at investigating in a common benchmarking scheme the core mechanisms that control the information flow, with specific focus on backpropagation and layerwise training.

\section*{Acknowledgments}
The author wishes to thank Giovanni Martines (former Professor of Electronics, University of Cagliari) and Andrea Manconi (Researcher at the Italian National Research Council) for their insightful discussions and valuable ideas that helped to shape this article.

\bibliography{bibliography} \bibliographystyle{plain} 

\end{document}